\documentclass[useAMS,usenatbib]{mn2e}

\pdfoutput=1

\usepackage{rotating,aas_macros,multirow}

\newcommand{\nh}{$N_{\mathrm{H}}$}
\newcommand{\ariel}{\textsl{Ariel V}}
\newcommand{\heao}{\textsl{HEAO I}}
\newcommand{\erg}{ \rm \, erg \, s^{-1} cm^{-2}}

\def\ergs{\ensuremath{\rm{\, erg \,s^{-1} }}}
\def\msun{\ensuremath{M_{\odot}}}
\def\mdot{\ensuremath{\dot{M}}}

\def\rxte{\textit{RXTE}}
\def\inte{\textit{INTEGRAL}}
\def\swift{\textsl{Swift}}
\def\xmm{\textsl{XMM/Newton}}
\def\chan{\textsl{Chandra}}
\def\igr{IGR~J17464$-$3213}
\def\h17{H1743$-$322}

\title[\h17: Radio/X-ray correlation]{Radiatively efficient accreting black holes in the hard state: the case study of \h17}
\author[M. Coriat et al.]{M. Coriat,$^{1}$\thanks{Now at School of Physics and Astronomy, University of Southampton, Southampton, SO17 1BJ, UK. E-mail:
m.coriat@soton.ac.uk} S. Corbel,$^{1,2}$ L. Prat,$^{1}$ J. C. A. Miller-Jones,$^{3,4}$ D. Cseh,$^{1}$   
\newauthor A. K. Tzioumis,$^{5}$ C. Brocksopp,$^{6}$  J. Rodriguez,$^{1}$ R. P. Fender$^{7}$ and G. R. Sivakoff$^{8}$\\
$^{1}$Laboratoire AIM, CEA-IRFU/CNRS/Universit\'e Paris Diderot, CEA Saclay, F-91191 Gif-sur-Yvette, France\\
$^{2}$Institut Universitaire de France, 75005 Paris, France\\
$^{3}$NRAO Headquarters, 520 Edgemont Road, Charlottesville, VA 22903, USA\\
$^{4}$International Centre for Radio Astronomy Research - Curtin University, GPO Box U1987, Perth, WA 6845, Australia\\
$^{5}$Australia Telescope National Facility, CSIRO, P.O. Box 76, Epping, NSW 1710, Australia\\
$^{6}$Mullard Space Science Laboratory, University College London, Holmbury St. Mary, Dorking, Surrey RH5 6NT, UK\\
$^{7}$School of Physics and Astronomy, University of Southampton, Highfield, Southampton, SO17 1BJ, UK\\
$^{8}$Department of Astronomy, University of Virgina, P.O. Box 400325, Charlottesville, VA 22904-4325, USA}

\begin{document}

\date{\today}

\pagerange{\pageref{firstpage}--\pageref{lastpage}} \pubyear{2011}

\maketitle

\label{firstpage}

\begin{abstract}
In recent years, much effort has been devoted to unraveling the connection between the accretion flow and the jets in accreting compact objects. In the present work, we report new constraints on these issues, through the long term study of the radio and X-ray behaviour of the black hole candidate \h17. This source is known to be one of the `outliers' of the universal radio/X-ray correlation, i.e. a group of accreting stellar-mass black holes displaying fainter radio emission for a given X-ray luminosity than expected from the correlation. Our study shows that the radio and X-ray emission of \h17 are strongly correlated at high luminosity in the hard spectral state. However, this correlation is unusually steep for a black hole X-ray binary: $b\sim 1.4$ (with $L_{\rm Radio} \propto L_{X}^{b}$). Below a critical luminosity, the correlation becomes shallower until it rejoins the standard correlation with $b \sim 0.6$. Based on these results, we first show that the steep correlation can be explained if the inner accretion flow is radiatively efficient during the hard state, in contrast to what is usually assumed for black hole X-ray binaries in this spectral state. The transition between the steep and the standard correlation would therefore reflect a change from a radiatively efficient to a radiatively inefficient accretion flow. Finally, we investigate the possibility that the discrepancy between `outliers' and `standard' black holes arises from the outflow properties rather than from the accretion flow.

\end{abstract}


\section{Introduction}

Black hole X-ray binaries (BHXBs) are binary systems consisting of a black hole primary in orbit with a less evolved companion star. These systems spend most of their time in a faint quiescent state, being barely detectable at almost all wavelengths. They may undergo sudden and bright few-month-long X-ray outbursts with typical recurrence periods of many years \citep{tanaka96}. The picture commonly accepted to explain the emission of such objects involves an optically thick and geometrically thin accretion disc, mostly emitting at typical energies of $\sim$1~keV. This region is probably surrounded by a corona of hot plasma, where UV and soft X-ray photons originating from the disc undergo inverse Compton scattering, producing a power law spectrum in the hard X-ray band. In addition to this `X-ray picture', BHXBs are also characterised by the intermittent presence of relativistic outflows. This ejected material is mainly detected at radio wavelengths \citep[see e.g.,][]{hjellming81,mirabel94,fender06} though it can sometimes dominate the low-frequency emission up to the near infrared \citep{corbel01,jain01,corbel02,homan05,russell06,coriat09}.  These jets are undoubtedly coupled to the accretion flow, although the nature of this connection is still unclear. 

Several spectral states have been identified based on the relative strengths and properties of the different X-ray emitting components (see e.g., \citealt{mcclintock06,homan05a}). The two main spectral states are the soft state, dominated by thermal emission from the accretion disc, and the hard state, dominated by emission from the corona. Various instances (hard or soft) of the intermediate state have also been defined to describe the transition phases between the two main states. During these phases the X-ray spectra usually display hardnesses in between those of the hard and the soft state as a result of comparable contributions to the emission from the disc and the corona. These spectral characteristics are also coupled to different levels of X-ray variability \citep[see e.g.,][for a review]{van-der-klis06, belloni10}. 

The radio emission in the hard state is usually characterized by a flat or slightly inverted spectrum ($S_{\nu} \propto \nu^{\alpha}$ with $\alpha \sim 0$). This is interpreted as self-absorbed synchrotron emission from steady, collimated, compact jets, in analogy with those observed in active galactic nuclei \citep{blandford79, hjellming88}. During the soft state these compact jets are thought to be quenched  \citep{fender99,corbel00} and any radio emission, if present, is attributed to residual optically thin synchrotron emission from transient ejecta (\citealt{corbel04}; \citealt*{fender04}).

Observations of several sources have provided evidence that a strong connection exists between radio and X-ray emission during the hard state (\citealt{hannikainen98,corbel00,corbel03}; \citealt*{gallo03}). This connection takes the form of a non-linear flux correlation, $F_{\rm Rad} \propto F_{\rm X}^b$, where $F_{\rm Rad}$ is the radio flux density, $F_{\rm X}$ is the X-ray flux and $b \sim 0.5-0.7$. It was subsequently shown that this same correlation also holds between optical-infrared (OIR) and X-ray fluxes \citep{homan05,russell06,russell07,coriat09}. These correlations indicate that the compact jets are strongly connected with the accretion flow (disk and/or corona), and possibly that their emission (synchrotron and/or inverse Compton) can comprise a significant contribution to the observed high-energy flux \citep[see e.g.,][]{markoff01,markoff03,markoff05,rodriguez08a,russell10}. This radio/X-ray correlation, initially established for the source GX 339-4, has been extended to other galactic black holes \citep*[mainly V404 Cyg;][]{gallo03,corbel08} and even active galactic nuclei \citep*{merloni03,falcke04,kording06a}. \citet{migliari06} also showed that a similar correlation exists for neutron star X-ray binaries (NSXBs), but with a steeper correlation coefficient ($b \sim 1.4$) and fainter radio emission for a given X-ray luminosity than seen in black holes \citep{fender00,fender01a,muno05}.

However, in the following years, a few galactic black hole candidates (BHCs) were found to lie well outside the scatter of the original radio/X-ray correlation (e.g., XTE J1720-318, \citealt{brocksopp05}; XTE J1650-500, \citealt{corbel04}; IGR J17497$-$2821, \citealt{rodriguez07}; Swift J1753.5-0127, \citealt{cadolle-bel07}, \citealt{soleri10}), thus either increasing its scatter or challenging the universality of the correlation itself. Some of them could also be false identifications of black holes. For a given X-ray luminosity, these outliers show a radio luminosity fainter than expected from the correlation (thus are sometimes dubbed `radio-quiet' BHCs). However, for most of these outliers, there are no radio measurements available at low X-ray luminosities. Therefore we do not know whether they remain underluminous in the radio band at low accretion rates. The current lack of data also precludes a precise measurement of the slope of the correlation (if any) for the outliers. It is therefore unclear whether they follow a correlation similar to the `standard' BHXBs but with a lower normalisation or whether their inflow/outflow connection is intrinsically different. Moreover, we do not know if their behaviour is recurrent over several outbursts. These are some of issues that we address in this work.

Note that in the following, we will use the term `outliers' rather than `radio-quiet BHCs' to describe these sources. As we will show, they could be considered `X-ray loud' as well. However, it should be borne in mind that the term `outliers' might not be appropriate either. Indeed, given the increasing number of these sources, the `outliers' could in fact turn out to be the norm.

\subsection*{\h17}

The X-ray transient \h17 was discovered with the \ariel\ and \heao\ satellites by \citet{kaluzienski77} during a bright outburst in 1977. In 2003, another bright outburst was first detected with the International Gamma-ray Astrophysics Laboratory (\inte). The source was initially dubbed \igr, before it was identified as \h17 \citep{markwardt03}. This outburst was extensively studied at all wavelengths \citep[see e.g.,][]{parmar03,joinet05,homan05b,capitanio05,lutovinov05, miller06a,kalemci06,mcclintock09}. It was shown in particular that the spectral and timing features of \h17 were similar to those of other, dynamically confirmed, black-hole X-ray transients \citep{mcclintock09}. It was thus classified as a BHC.

During the return to quiescence following this outburst, \citet{corbel05} reported the detection of large scale, synchrotron-emitting jets moving away from the central source. These jets were detected at both radio and X-ray wavelengths as a consequence of the interaction between the ejected plasma and the interstellar medium (ISM). Using the observed proper motions of the X-ray jets, these authors also derived an upper limit to the source distance of $10.4 \pm 2.9$ kpc. Given its location ($l = 357.255$ and $b = -1.83$) in the direction of the Galactic bulge, and a rather high column density, this upper limit is consistent with a Galactic Center distance for \h17. In the following, we will therefore assume a distance of 8 kpc.

\begin{figure*}
\centering
\begin{minipage}{1.0\textwidth}
    \includegraphics[width=0.5\textwidth]{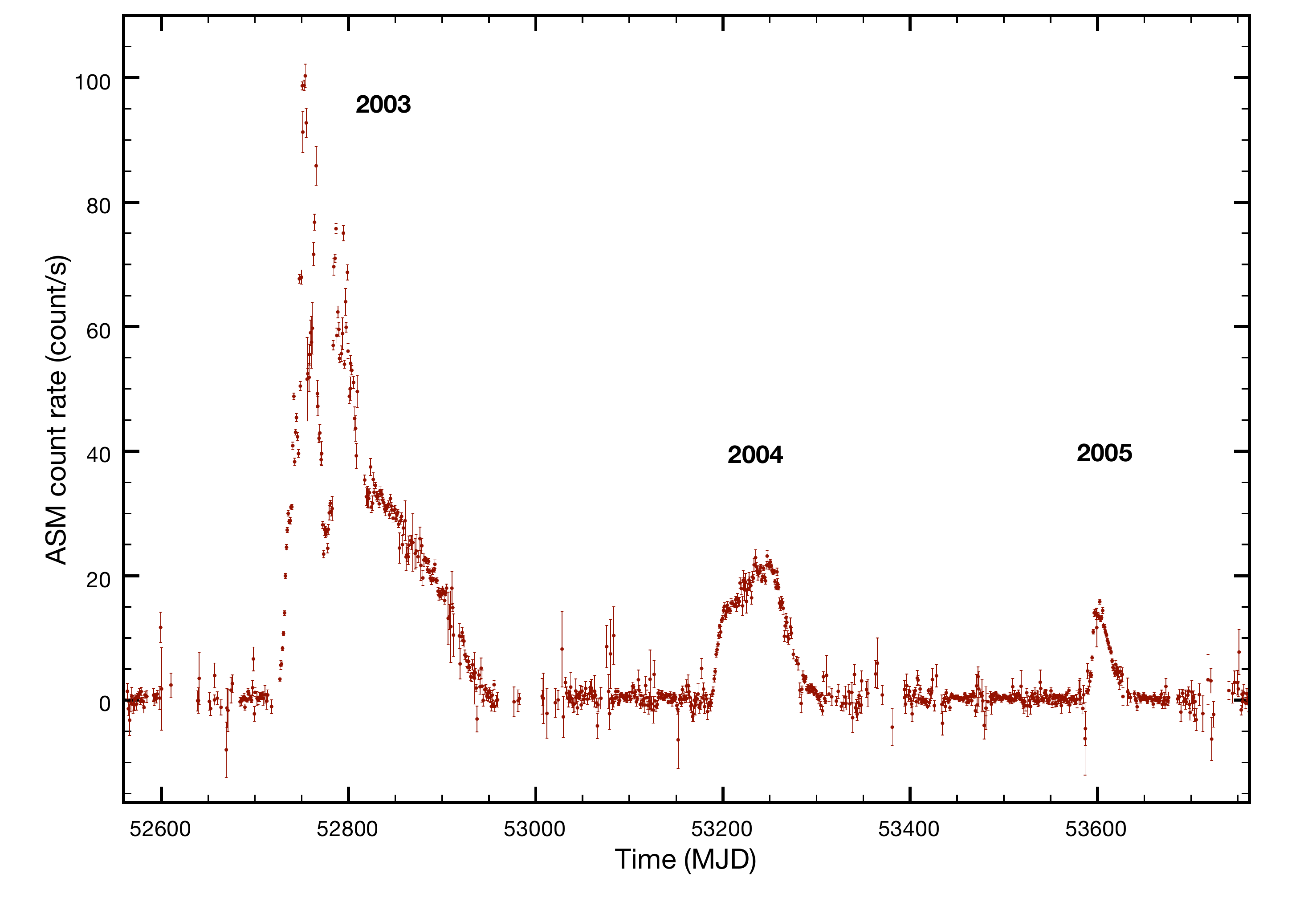} 
        \includegraphics[width=0.5\textwidth]{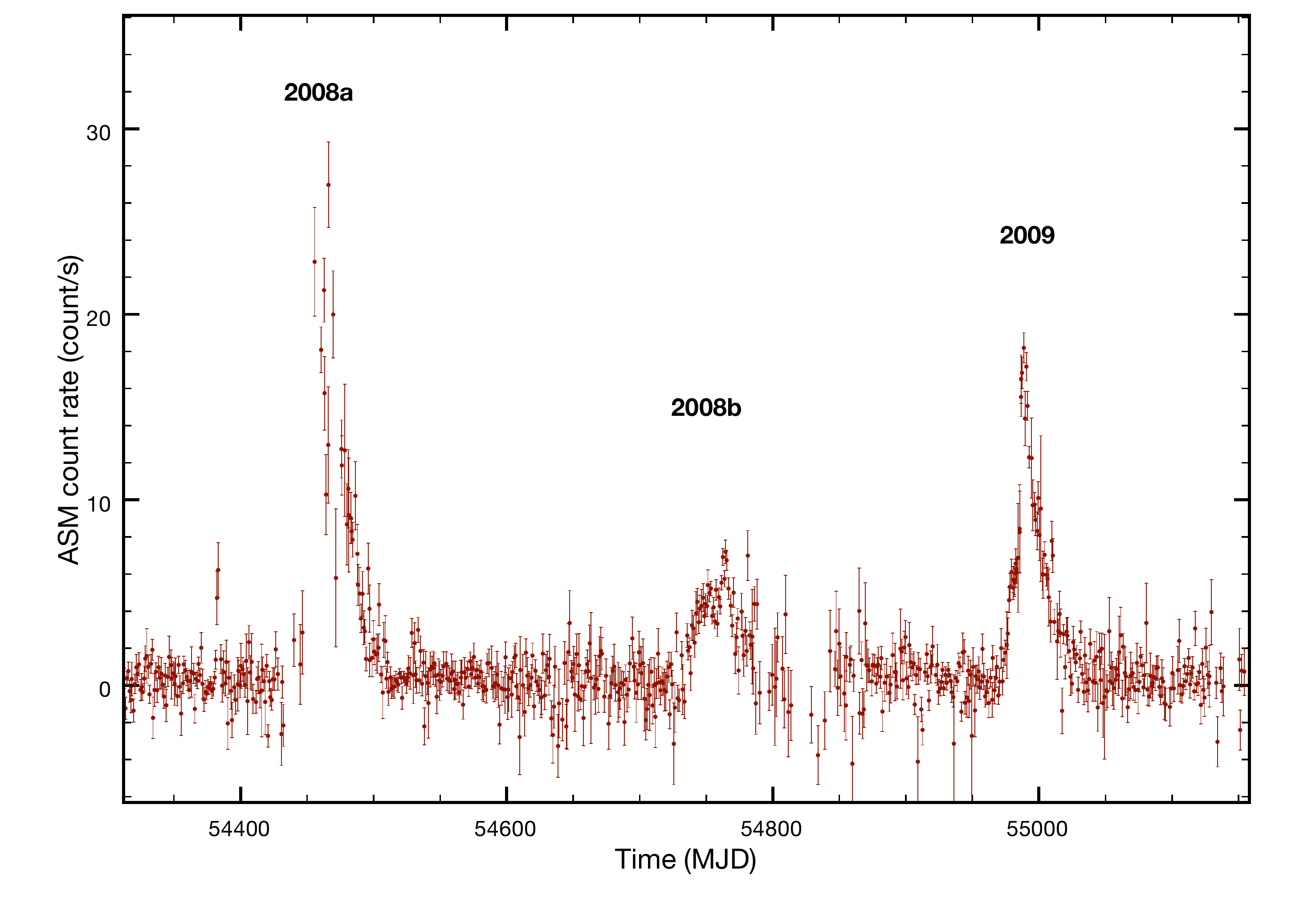} 
           \caption{\rxte/ASM light curve of \h17 between 2003 and 2009. We note a major outburst in 2003 followed by five minor activity periods between 2004 and 2009. Note the different scaling between the two plots. The 2010 light curve is not presented due to the very low number of counts detected by the ASM during this short outburst.}
  \label{asm}
              \end{minipage}
  \end{figure*}

The 2003 outburst was followed by weaker outbursts (see Fig.~\ref{asm}) in 2004 \citep{swank04} and 2005 \citep{rupen05} which were poorly sampled at both X-ray and radio wavelengths. Therefore, no detailed studies of these two phases have been carried out to date. Further outbursts were observed in the first months of 2008 \citep[2008a in the following;][]{kalemci08,jonker10} and in 2008 September--November \citep[2008b in the following;][]{corbel08a}. During the outburst decay of the 2008a outburst, \citet{jonker10} reported a radio/X-ray correlation slope of $b = 0.18 \pm 0.01$. The authors also found that \h17 lies well below the `universal' radio/X-ray correlation making of \h17 another outlier. The weak 2008b outburst was classified as `failed'. The source made a short cycle between the hard and the hard intermediate state but never reached the soft state \citep{prat09,capitanio09}. The source entered another outburst phase in 2009 \citep{krimm09} and also in early 2010 \citep{yamaoka09}. In 2009, the system followed the canonical evolution through all the characteristic states \citep{motta10}. The variation of the flux associated with the two main spectral components (i.e. disc and power law) allowed \citet{motta10} to set a lower limit to the orbital inclination of the system of $\geqslant 43\degr$.

In this work, we focus on the long-term study of the radio/X-ray correlation over the 6 outbursts mentioned above. We aim to investigate in detail the accretion-ejection coupling in this system in the global context of the `outliers' of the radio/X-ray correlation.
The sequence of observations and data reduction processes are described in Section 2. In Section 3, we present the analysis of the radio/X-ray correlation, the selection process that we applied to isolate and study the connection between the compact jets and the inner accretion flow, and finally a comparison with other black hole and neutron star X-ray binaries. These results are then discussed in Section 4, in which we investigate several possible interpretations. Our conclusions are summarised in Section 5.

\section{Observations}

\begin{figure*}
\centering
\begin{minipage}{1.0\textwidth}
    \includegraphics[width=0.5\textwidth]{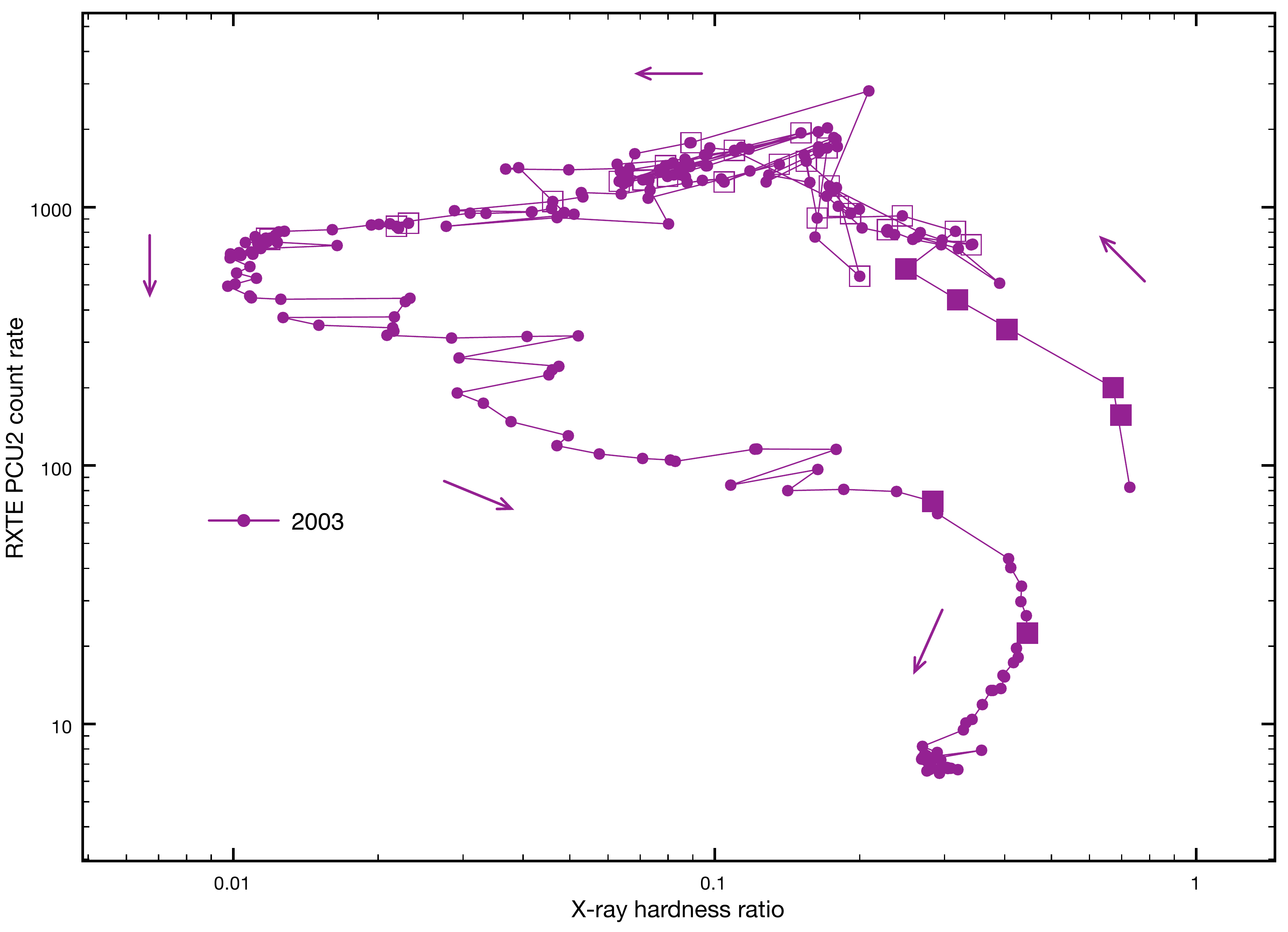} 
        \includegraphics[width=0.5\textwidth]{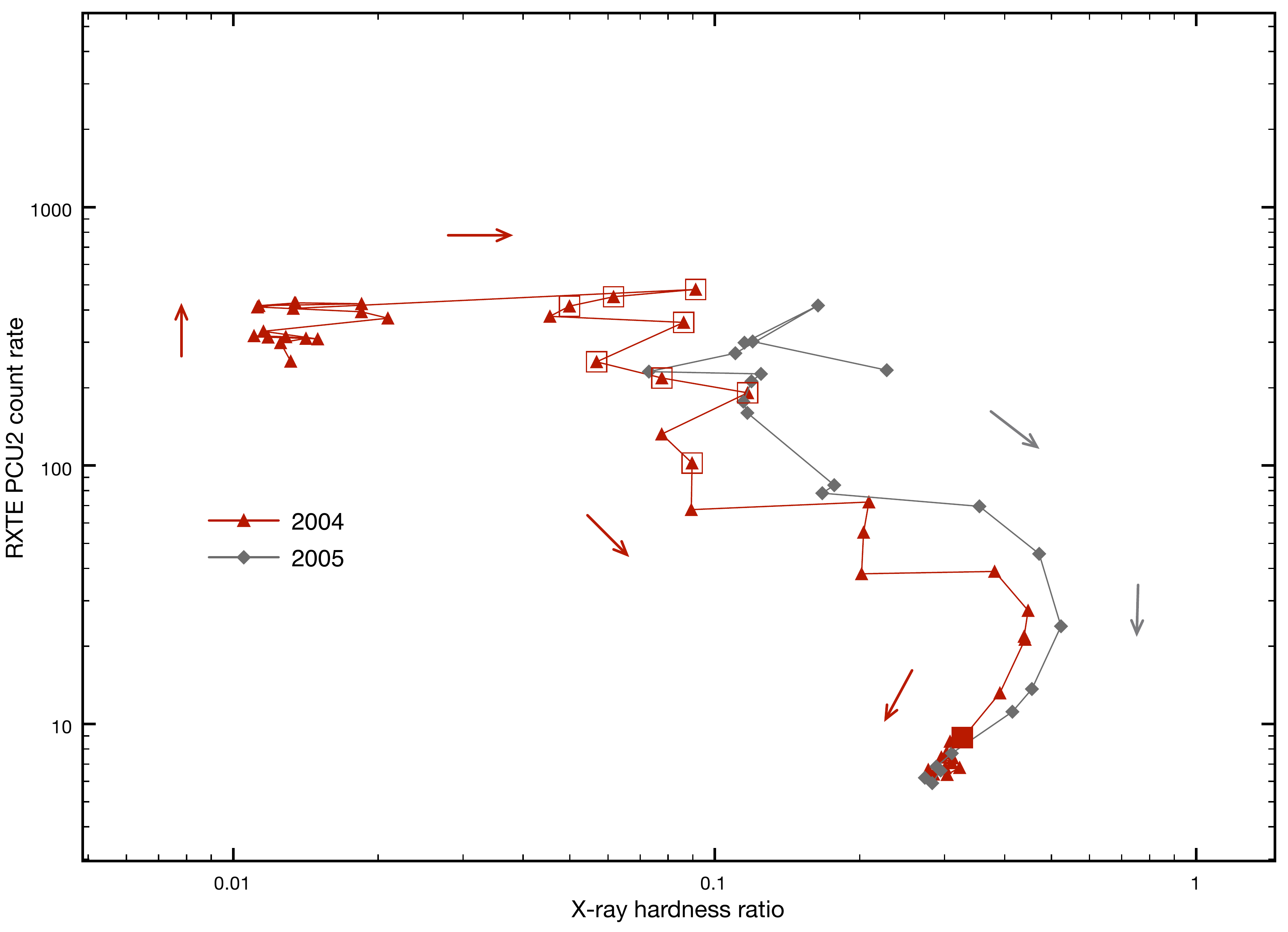} 
              \end{minipage}
                \includegraphics[width=0.7\textwidth]{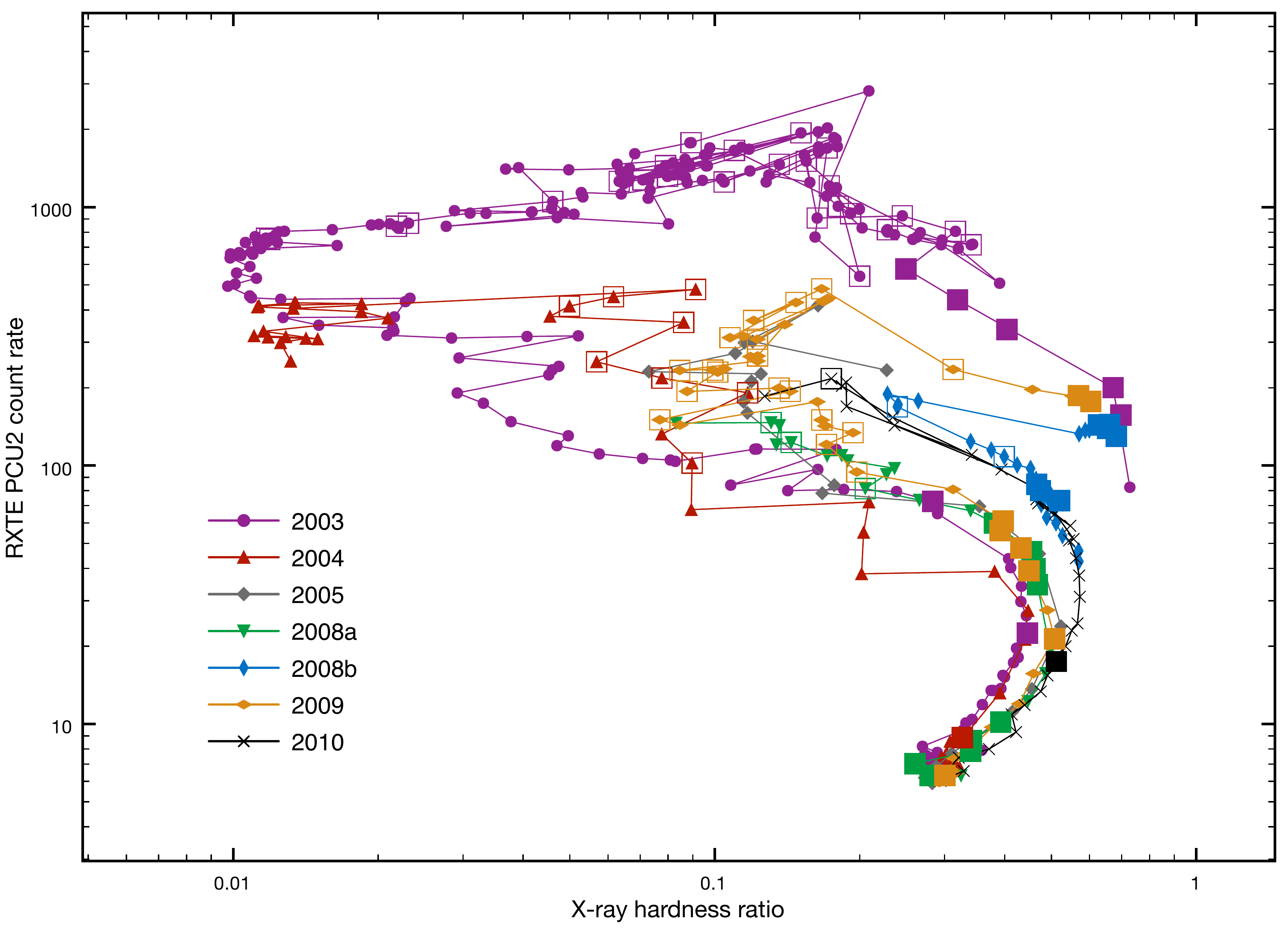} 
                \begin{minipage}{1.0\textwidth}
	        \includegraphics[width=0.5\textwidth]{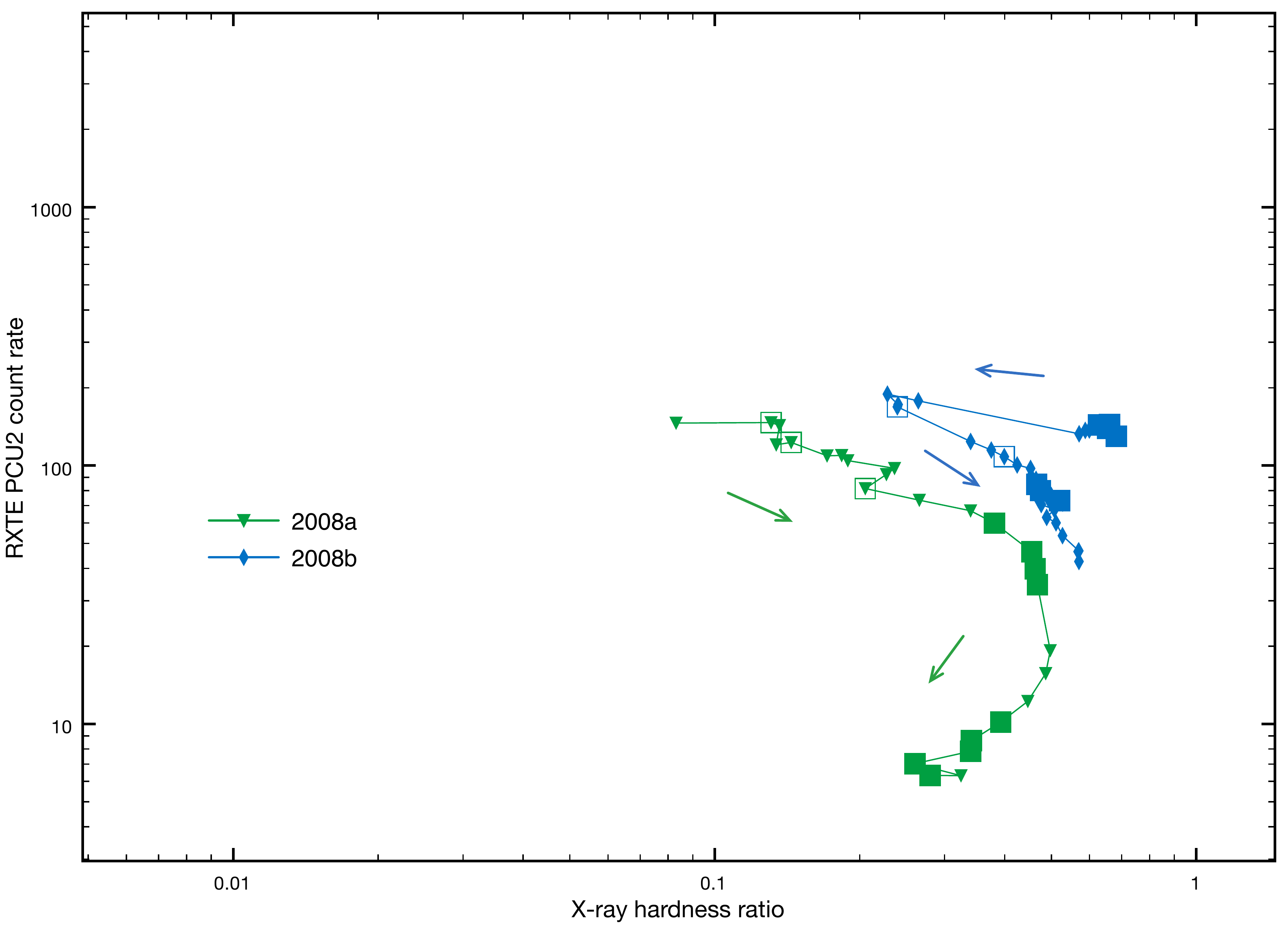} 
	                \includegraphics[width=0.5\textwidth]{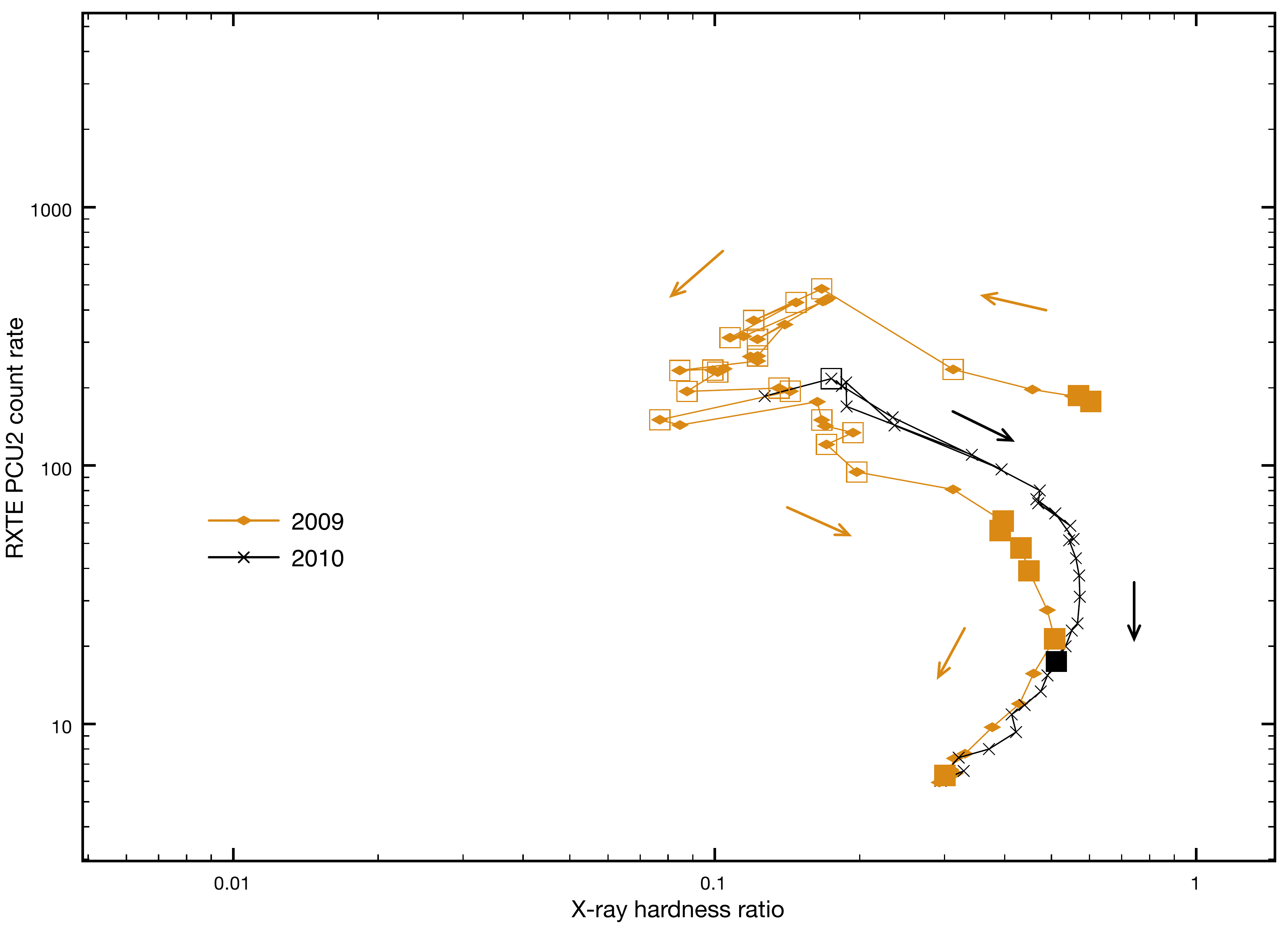} 
           \caption{Hardness-intensity diagrams (HIDs) of \h17 from 2003 to 2010. Squares (open and filled) indicate the radio detections, plotted on top of the HIDs at the location of the nearest RXTE observation. Filled squares indicate the data selected for the radio/X-ray diagram on Fig. \ref{filtre}. The radio and X-ray fluxes corresponding to this selection are detailed in Table \ref{table-correl}. Arrows indicate the temporal evolution during the outbursts.  X-ray observations are indicated by purple circles for the outburst in 2003, red triangles for 2004, grey diamonds for 2005, inverted green triangles for 2008a, blue diamonds for 2008b, orange lozenges for 2009, and black crosses for 2010.}
  \label{hid}
              \end{minipage}
  \end{figure*}

\subsection{X-ray }

\subsubsection{\textit{RXTE}: Data reduction and spectral analysis}

 We analysed all publicly available observations of \h17 in the \rxte\ archive taken between 2003 January 1 and 2010 February 13. 
The data were reduced using the {\tt HEASOFT} software package v6.8, following the standard steps described in the \rxte\ cookbook\footnote{http://heasarc.gsfc.nasa.gov/docs/xte/data\_analysis.html}. Spectra were extracted from the Proportional Counter Array (PCA; \citealt{jahoda06}) in the 3--25 keV range. We only used the top layer of the Proportional Counter Unit (PCU) 2 as it is the only operational unit across all observations and is the best-calibrated detector out of the five PCUs. Systematic errors of 0.5 per cent were added to all channels. In the 20--150 keV range, we used data from the High Energy Timing Experiment (HEXTE), which we reduced following standard steps. From December 2005, due to problems in the rocking motion of Cluster A, we extracted spectra from Cluster B only. Due to the low count rate in the HEXTE data in most of the observations, all channels were rebinned by a factor of 4.

In addition, we constructed Hardness Intensity Diagrams (HIDs) from PCA data. These data were extracted from PCU2 (all layers) and corrected for background. Averaged count rates were extracted in two bands: (standard 2) channels $2-10$ and $19-40$, corresponding to $2.5-6.1$ and $9.4-18.5$ keV, respectively. The hardness ratio was defined as the ratio of the flux in the second band to that in the first band, and the intensity was calculated as the sum of the fluxes in both bands.

We performed a simultaneous fit to the PCA and HEXTE spectra in {\tt XSPEC V12.5.1n}, using a floating normalisation constant to allow for cross-calibration uncertainties.
The main objective of the X-ray spectral analysis was to obtain a correct estimation of the unabsorbed flux in the 3-9 keV band. Consequently, we used simple models
to reproduce the spectra and achieve statistically acceptable fits (i.e., a reduced $\chi^2 < 2$).
We used a power-law ({\tt powerlaw}) and an absorption component ({\tt phabs}) as a starting model.
When required by an F-test, we added a multi-temperature disc blackbody ({\tt diskbb}) and/or a high energy cutoff ({\tt highecut}). Eventually, when the residuals indicated the presence of reflection features, we used a Gaussian emission line ({\tt gaussian} constrained in energy between 6 and 7 keV) and smeared absorption edge ({\tt smedge}, constrained in energy between 7 and 9 keV).
The hydrogen column density was fixed at the value obtained  by \citet{prat09}\footnote{Note that several estimates of the hydrogen column density are found in the literature. The values range from $1.6  \times 10^{22} \; {\rm cm}^{-2}$ \citep[e.g.,][]{capitanio09} to $2.3  \times 10^{22} \; {\rm cm}^{-2}$ \citep[e.g.,][]{miller06a}. However, within this range, the precise value of the \nh\ and its possible variation during the outburst has little influence on the unabsorbed flux above 3 keV.  We thus used the intermediate value found by \citet{prat09}.} using  \swift\ and \xmm\  observations i.e. \nh$ =1.8 \pm 0.2 \times 10^{22} \; {\rm cm}^{-2}$.
At low count rates, when \h17 was not significantly detected by HEXTE, fits were made to the PCA spectrum only.  
We finally obtained an average reduced $\chi^2$ of 1.04 with a minimum of 0.64 and a maximum of 1.35. Unabsorbed fluxes were then estimated in the 3--9 keV energy ranges, according to the PCA normalisation.

Due to the location of the source close to the Galactic plane, the Galactic ridge emission starts to significantly contaminate the estimated 3--9 keV PCA flux below $\sim 10^{-10} \erg$.
\citet{kalemci06} determined a 3--25 keV unabsorbed flux from the ridge emission of $1.08 \times 10^{-10} \erg$, based on the analysis of nine observations in 2004 (MJD 53021--53055). 
We analysed the same data set to estimate the ridge emission in the 3--9 keV band and found an unabsorbed flux of $(6.0 \pm 0.6) \times 10^{-11} \erg$. Consequently, we subtracted this value from all 3--9 keV PCA fluxes. To check whether a simple subtraction was appropriate to correct the measured source flux for the contamination from the ridge emission, we combined the spectra of the 9 observations mentioned above to obtain a typical ridge spectrum. We then used this as an additional background spectrum for several on-source observations, where the ridge emission made a significantly contribution.  Finally, we compared the 3--9 keV fluxes obtained using this method with those obtained via simple flux subtraction. There was no significant difference within the error bars.

\subsubsection{Other X-ray satellites}
\label{sec:jonker}

\citet{jonker10} studied the decay of the 2008a outburst using \chan\ and \swift\ X-ray data simultaneous with radio observations from the Very Large Array (VLA). 
 Since they provide important constraints on the correlation at low luminosity, we included in our study the X-ray fluxes published in their paper. 
 We converted the unabsorbed 0.5--10 keV fluxes into absorbed 3--9 keV fluxes with the \textit{WebPimms} tool\footnote{http://heasarc.gsfc.nasa.gov/Tools/w3pimms.html} using the \nh\ and photon index provided by the authors.
For consistency with the \rxte\ data, we then calculated the unabsorbed 3--9 keV fluxes using  \nh$=1.8 \times 10^{22} {\rm cm}^{-2}$.

\subsubsection{X-ray state classification}

Since the definition and nature of the X-ray states is still debated, in the following, we will adopt a simplified classification adapted to the purpose of this work. Our aim is to understand the nature of the connection between the corona and the compact jets for the outliers of the radio/X-ray correlation. Therefore, we will be mainly interested in phases where the compact jets are present and where the X-ray emission in the 3--9 keV band is dominated by the power-law emitting component. Consequently, we will define only three states: hard, soft and intermediate. To be classified as hard state, we require that the power-law component dominates the X-ray spectrum in the sense that an accretion disc component was not required (by an F-test) to correctly fit the data above 3 keV. We also require a power-law photon index $\Gamma<2$. We define the soft state by a power-law photon index $\Gamma>2$ and a disk flux comprising $>75$ per cent of the 3-9 keV flux. All observations that do not correspond to either of these criteria are classified as intermediate states.

\subsection{Radio}

\subsubsection{ATCA}
Between 2003 April 24 (MJD 52753) and 2010 February 13 (MJD 55240), we performed a total of 38 observations of \h17\ with the Australia
Telescope Compact Array (ATCA). From 2009 April, the observations were carried out using the Compact Array Broadband Backend (CABB).
This upgrade has provided a new broadband backend system for the ATCA, increasing the maximum bandwidth from 128\,MHz to 2\,GHz. Each observation
was conducted simultaneously in two different frequency bands, with central frequencies of 4.8 GHz and 8.64 GHz (5.5 GHz and
9 GHz respectively following the CABB upgrade). Various array configurations were used during these observations.

The ATCA has orthogonal linearly polarised feeds and full Stokes parameters (I, Q, U, V) are recorded at each frequency.  We used PKS 1934--638
 for absolute flux and bandpass calibration, and PMN 1729--37 to calibrate
the antenna gains and phases as a function of time. We determined the polarisation leakages using either the primary or the secondary calibrator, 
depending on the parallactic angle coverage of the secondary. 
Imaging was carried out using a combination
of multi-frequency \citep{sault94} clean and standard clean algorithms.
The editing, calibration, Fourier transformation, deconvolution and image analysis were carried out with the Multichannel Image Reconstruction, 
Image Analysis and Display (MIRIAD) software \citep*{sault95}.

\subsubsection{VLA}

\h17 has also been regularly observed between 2003 and 2010 with the VLA. To extend our data set,
we made use of the radio flux densities at 4.86 GHz and 8.46 GHz published in \citet{mcclintock09} for the 2003 outburst, in 
 \citet{rupen04,rupen05} for the 2004 and 2005 outbursts and in \citet{rupen08,rupen08a} and \citet{jonker10} for the 2008a outbursts. We collected 
 a total of 68 VLA pointings. All VLA data are summarised in the aforementioned references in which data reduction and analysis are detailed. 
 In addition, we retrieved unpublished archival data of the 2004 outburst (PI: Rupen) from the National Radio Astronomy Observatory (NRAO) database.
All data were reduced using standard procedures within the NRAO {\tt AIPS} software package, using 3C286 as our primary calibrator, and J1744-3116 as the secondary calibrator.

For the 2009 outburst, we triggered VLA observations of \h17 after detection of an X-ray flare by Swift/BAT on 2009 May 26 \citep{krimm09}.  
On 2009 May 27, we detected unresolved radio emission at 8.4 GHz and triggered a monitoring campaign to cover the outburst of the source from the rising hard state through the decay back to quiescence.  
Our final observation was taken on 2009 August 6.  Observations were made in dual circular polarisation in each of two contiguous intermediate frequency pairs, giving a total bandwidth of 100 MHz per polarization.  
We observed primarily at 8.4 and 4.8 GHz, but also at 1.4 GHz when the source flux density was predicted to be above 0.3 mJy, and at 22.4 GHz for two epochs at the peak of the flare, although the source was not detected in either observation at this frequency.  The array was in its intermediate CnB and C configurations throughout the duration of our observing campaign.  

\subsection{Simultaneity}
\label{sec:simult}
For the vast majority of the radio data, we found quasi-simultaneous ($\Delta t \leqslant$ 1 day) \rxte\ observations. Otherwise, we interpolated the X-ray flux from a polynomial fit to the PCA lightcurve. 
We estimated that the uncertainty introduced by this method should be less than 15 per cent as the flux evolution was found to be smooth in all cases.
When the missing flux was not framed by at least two X-ray pointings, we converted the \rxte/All Sky Monitor (ASM) count rate into  3--9 keV unabsorbed flux with \textit{WebPimms}, using the spectral parameters of the nearest X-ray observation.

\section{Radio/X-ray correlation}

\subsection{Overview}

\begin{figure*}
\centering
\begin{minipage}{1.0\textwidth}
    \includegraphics[width=0.5\textwidth]{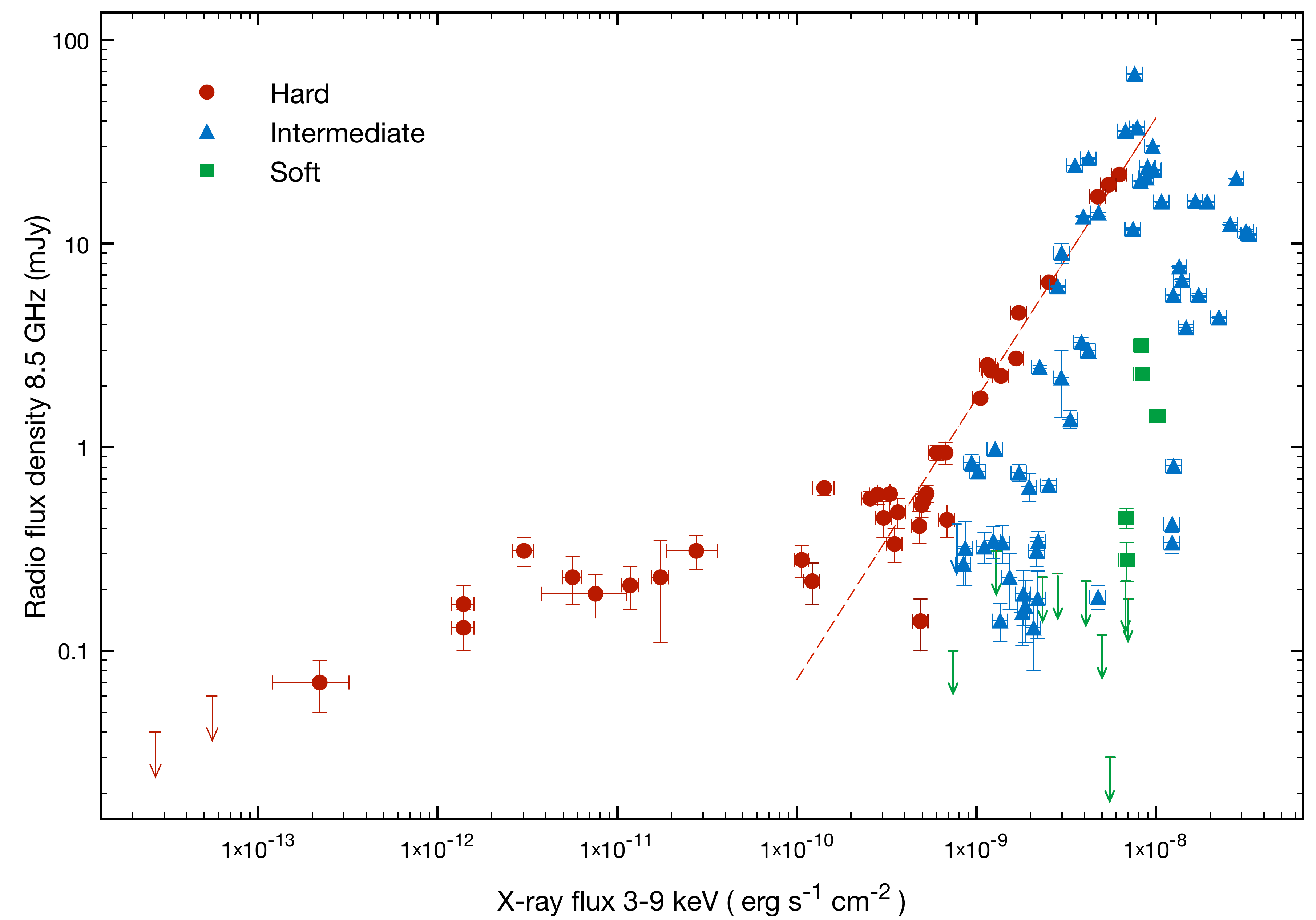} 
        \includegraphics[width=0.5\textwidth]{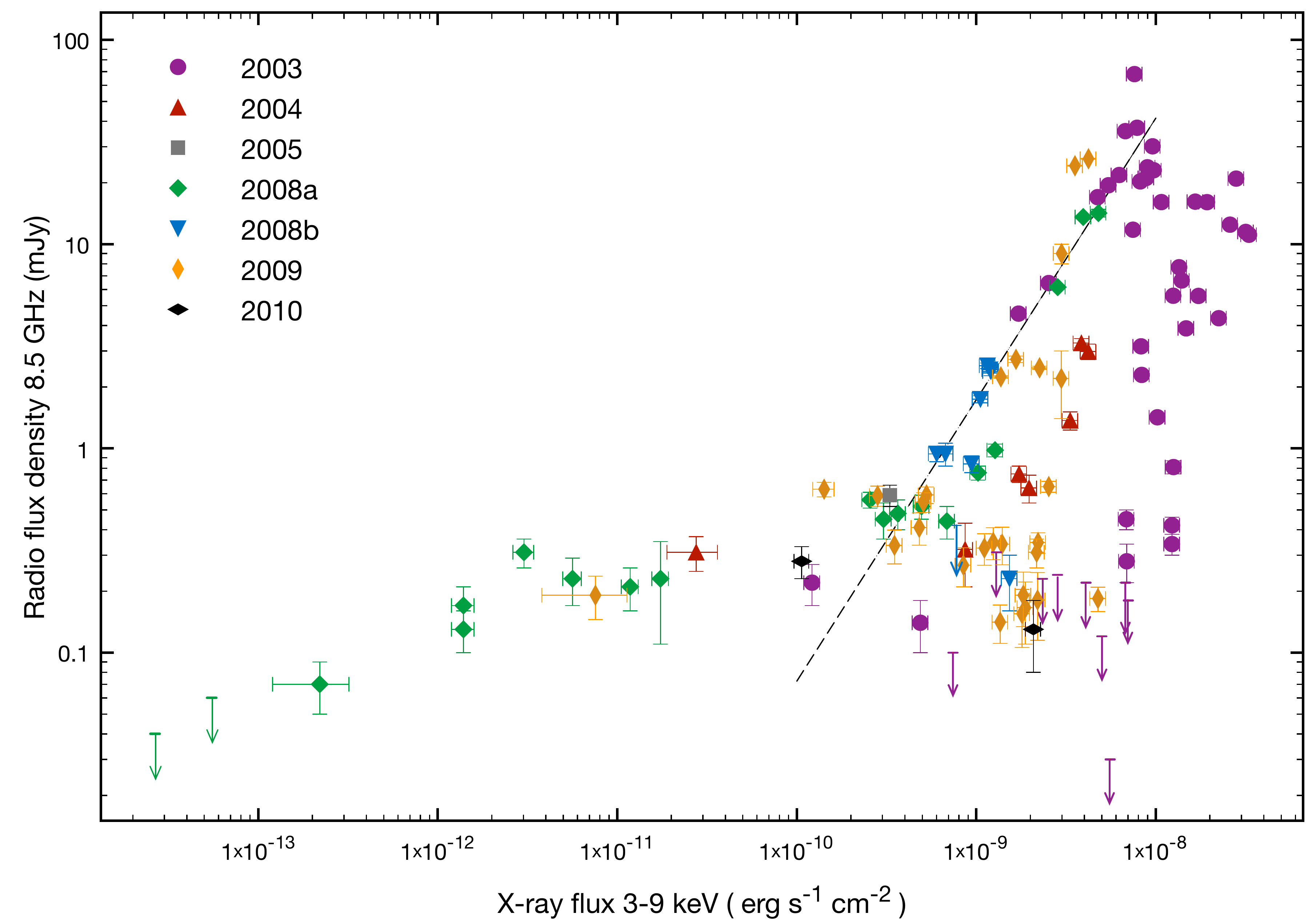} 
           \caption{Quasi-simultaneous 8.5 GHz radio flux density versus 3-9 keV X-ray flux during the seven outbursts. \textit{(a) Left panel:} Data are colour coded according to X-ray state. Red circles, green squares and blue triangles indicate the hard, soft and intermediate state respectively. Dashed line indicates the power-law fit to the selected hard state data (see Section~\ref{sec:isolating}). \textit{(b) Right panel:} Same as left panel but with the colour code indicating the outburst, as labelled in the figure legend.}
  \label{RXtotal}
              \end{minipage}
  \end{figure*}

In a first approach to characterise the global radio-X-ray behaviour of \h17, we use the complete data set without restricting either the X-ray state or the origin of the radio emission (e.g., compact jet, discrete ejecta, interaction with the ISM). Fig. \ref{RXtotal} shows the 8.5 GHz radio flux density\footnote{We use an average frequency of 8.5 GHz for simplicity since the radio data come from the VLA (8.46 GHz) and the ATCA (8.64 GHz). Even for an optically thin spectrum with spectral index $\alpha=-0.6$, the effect of this simplification would be $<1.5$ per cent.} versus the  3--9 keV unabsorbed flux over the 2003, 2004, 2005, 2008a, 2008b, 2009 and 2010 outbursts. The left panel of Fig. \ref{RXtotal} shows the data grouped according to the X-ray state. The right panel show the data grouped by outburst.

\subsubsection*{Behaviour at high flux}

Above a 3--9 keV X-ray flux $\sim 2 \times 10^{-10} \erg$, the behaviour of \h17 seems compatible, on first inspection, with the radio/X-ray behaviour typically observed during an outburst of a BHXB
(for a detailed discussion on this unified picture see , e.g., \citealt*{fender09a} and references therein). The hard state shows correlated radio and X-ray emissions over two orders of magnitude in radio flux density. On average, it is the most radio loud state for a given X-ray luminosity. During the intermediate state, the X-ray emission increases but is not correlated with the radio emission. 
For most of the radio observations during the intermediate state, the spectral index is indicative of optically thin synchrotron emission from transient ejecta ($\alpha \la -0.5$, where radio flux density, $S_{\nu}$ scales with frequency $\nu$ as $S_{\nu}\propto\nu^{\alpha}$). 
During the soft state, the radio emission is usually not detected as illustrated by the upper limits, however, we note several radio detections during the soft state of the 2003 outburst. This residual emission shows optically thin spectra and could originate from the interaction of the ejected matter with the environment, as was later observed on larger scales \citep{corbel05}.

\subsubsection*{Behaviour at low flux}

If we consider now the entire plot including the low flux data, we note several data points that clearly depart from the main hard state correlation below $\sim 2 \times 10^{-10} \erg$ in the 3--9\,keV X-ray band. Most of these points (green diamonds in Fig. \ref{RXtotal}b) belong to the decay phase of the 2008a outburst and were obtained using the \chan\  and \swift\  satellites along with the VLA \citep{jonker10}. The origin of the radio emission is unclear since most of the VLA observations were conducted at only one frequency. For three of them, however, upper limits at 1.4 GHz are available. The corresponding lower limits on the radio spectral indices ($\alpha \geqslant -0.58, -0.53 \; \rm{and} \, -0.60$) encompass both possibilities of optically thick and thin spectra. On the other hand, we note that the data from the declining hard state of the 2004 and 2009 outbursts (and possibly also 2003 and 2010) also deviate from the main correlation and seem to follow the same trend as the 2008a data. Moreover, their nearly flat radio spectra are consistent with a compact jet origin. This would suggest that this deviation is a significant evolution of the inflow - outflow connection when the source reaches low luminosities.

\subsubsection*{Jet quenching factor}

The drop in radio flux density during the hard to soft state transition is usually attributed to the quenching of the compact (core) jets. To estimate the level of this suppression, we can use the ratio between the highest radio flux attributed to compact jet emission from the initial hard state of an outburst, and the lowest radio upper limit obtained during the soft state of the same outburst. The $3 \sigma$ upper limit of 0.03 mJy obtained on MJD 52863 during the soft state of the 2003 outburst, provides a quenching factor of $\sim 700$, which is, as far as we know, the strongest constraint to date \citep{fender99,corbel01,corbel04} supporting the idea of jet suppression during the soft state.

\subsection{Isolating the compact jet - corona connection}
\label{sec:isolating}

\begin{figure}
	 \includegraphics[width=0.50\textwidth]{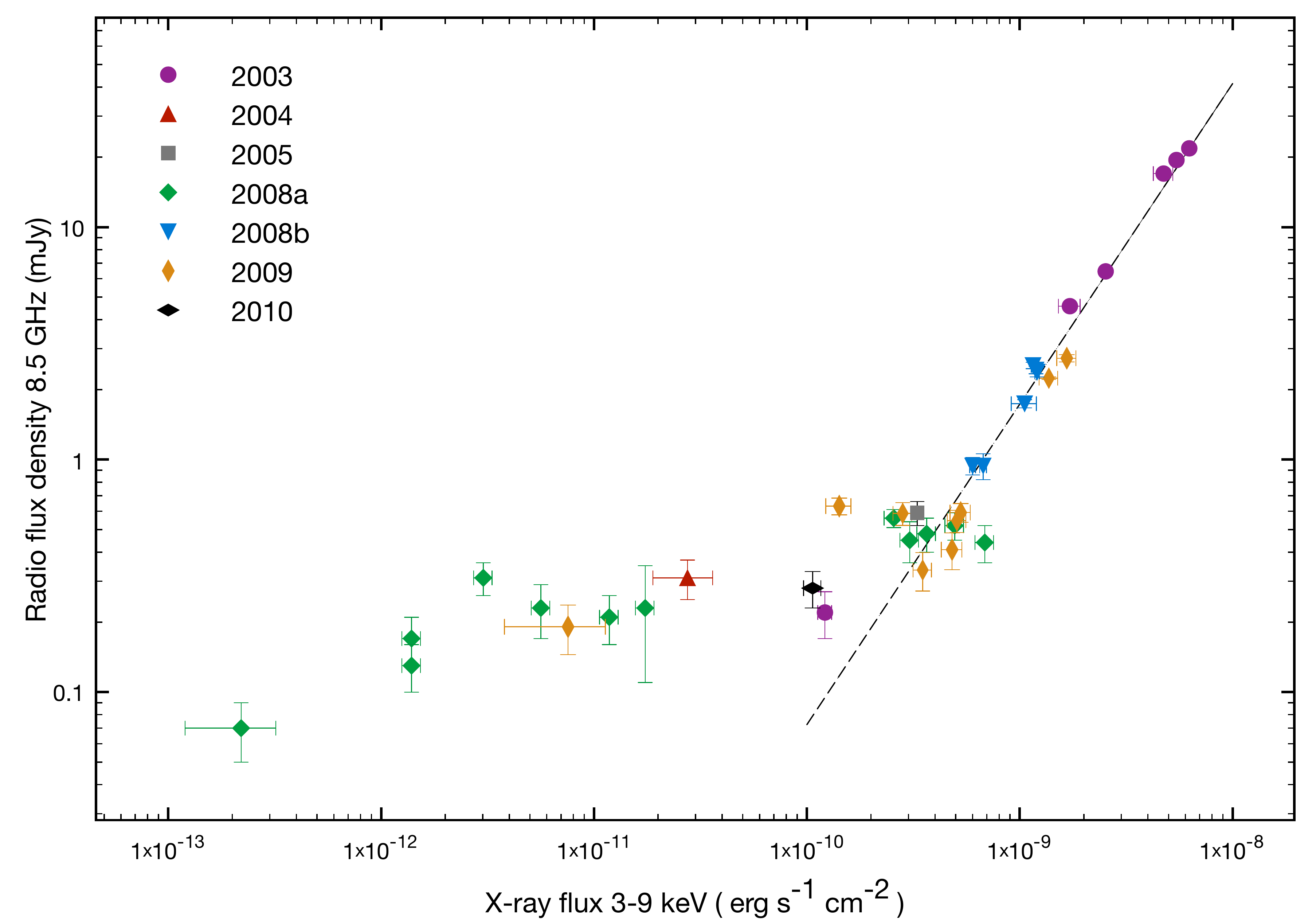} 
	\caption{Radio flux density at 8.5 GHz versus the unabsorbed 3--9 keV X-ray flux. This plot shows the data set 
	restrained to the `canonical' hard state where the radio emission can be attributed to partially self-absorbed synchrotron emission from the compact jets and where the X-ray spectra are dominated by the power-law component. Dashed line indicate the fit to the data above $2 \times 10^{-10} \erg$ with a function of the form $F_{\rm Rad} = k \, F_{\rm X}^b$, with $b = 1.38 \pm 0.03$ and $k = 4.43 \times 10^{12}$\,mJy\,erg$^{-1}$s\,cm$^2$.}
	\label{filtre}
\end{figure}

\begin{table*}
\caption{X-ray fluxes and radio flux densities selected for the correlation presented on Fig. \ref{filtre}. Unless otherwise stated, the X-ray data are from the \textit{RXTE-PCA} instrument. All \textit{RXTE} fluxes are corrected for the Galactic ridge emission. We also indicate the interferometer used for the radio observations. }
 \label{table-correl}
\begin{tabular}{lcccl}
\hline
\hline
Calendar date&MJD & X-ray 3-9 keV unabs. flux &Radio 8.5 GHz flux density&Notes\\
&& $(10^{-11} \erg)$ & (mJy)& \\
\hline
2003 March 30 &52728.59 &$ 172\pm20$ &$ 4.57\pm0.12$& \textit{VLA} (1,A)\\
2003 April 01&52730.55 & $253\pm1$ & $6.45\pm0.12 $& \textit{VLA} (1)\\
2003 April 03 &52732.55 & $474\pm50$ &$ 16.99\pm0.12$& \textit{VLA} (1,B)\\
2003 April 04 &52733.55 & $626\pm2 $& $21.81\pm0.13 $&\textit{VLA} (1)\\
2003 April 06 &52735.47 & $545\pm7$ &$ 19.43\pm0.16$ &\textit{VLA} (1)\\
2003 November 05 &52948.00 & $12.16\pm0.91$ & $0.22\pm0.05 $&\textit{VLA} (1)\\
&&&&\\
2004 November 01 &53310.87 & $2.75\pm0.86$ &$ 0.31\pm0.06$ &\textit{ATCA}\\
&&&&\\
2005 August 07 &53589.25 &$ 33.0\pm1.5$ & $0.59\pm0.07$& \textit{VLA} (2,A) \\
&&&&\\
2008 January 28 &54493.32 & $68.5\pm5.3$ &$ 0.44\pm0.09$ &\textit{ATCA}\\
2008 February 03 &54499.74 & $49.5\pm3.6$ &$ 0.52\pm0.07$&\textit{VLA} (3)\\
2008 February 05 &54501.64 & $36.5\pm4$ & $0.48\pm0.08$ &\textit{VLA} (3)\\
2008 February 06 &54502.56 & $30.424\pm3$ & $0.45\pm0.09$ &\textit{VLA} (3)\\
2008 February 09 &54505.67 & $25.60\pm3 $& $0.56\pm0.05$&\textit{VLA} (3)\\
2008 February 19 &54515.63 &$ 1.74\pm0.25 $& $0.23\pm0.12$ &\textit{Swift - VLA} (3,4)\\
2008 February 20 &54516.55 & $1.18\pm0.3 $& $0.21\pm0.05$ &\textit{Chandra - VLA} (3,4)\\
2008 February 23 &54519.69 & $0.56\pm0.06$ & $0.23\pm0.06$ &\textit{Chandra - VLA} (3,4)\\
2008 February 24 &54520.69 & $0.30\pm0.04$ & $0.31\pm0.05$ &\textit{Swift - VLA} (3,4)\\
2008 March 01 &54526.59 & $0.14\pm0.02$ & $0.17\pm0.04$&\textit{Chandra - VLA} (3,4)\\
2008 March 02 &54527.53 & $0.14\pm0.02$ & $0.13\pm0.03$ &\textit{Chandra - VLA} (3,4)\\
2008 March 08 &54533.56 & $0.022\pm0.01 $& $0.07\pm0.02$ &\textit{Chandra - VLA} (3,4)\\
&&&&\\
2008 October 05 &54744.21 & $105\pm14$ & $1.74\pm0.07$ &\textit{ATCA}\\
2008 October 08 &54747.44 & $115\pm2 $& $2.54\pm0.08$&\textit{ATCA} \\
2008 October 09 &54748.44 & $119\pm2 $& $2.43\pm0.09$ &\textit{ATCA}\\
2008 October 10 &54749.36 & $121\pm3 $& $2.38\pm0.11$&\textit{ATCA} \\
2008 November 04 &54774.43 & $67.3\pm2.5 $& $0.94\pm0.12$&\textit{ATCA} \\
2008 November 09 &54779.35 & $60.1\pm1.9 $& $0.94\pm0.08$ &\textit{ATCA}\\
&&&&\\
2009 May 27 &54978.38 & $137\pm14 $& $2.24\pm0.03$& \textit{VLA} (A)\\
2009 May 30 &54981.95 & $166\pm7$ & $2.73\pm0.10$&\textit{VLA} \\
2009 July 07 &55019.46 & $52.8\pm5.8$ & $0.592\pm0.055$ &\textit{VLA}\\
2009 July 08 &55020.30 & $50.7\pm5 $& $0.546\pm0.06$ &\textit{VLA} (B)\\
2009 July 09 &55021.42 & $48.1\pm5.3$&$ 0.41\pm0.074$ &\textit{VLA}\\
2009 July 12 &55024.28 & $35.0\pm3 $&$ 0.335\pm0.063$ &\textit{VLA} (B)\\
2009 July 13 &55025.28 & $28.2\pm1.9 $&$ 0.587\pm0.066$&\textit{VLA} \\
2009 July 19 &55031.29 & $14.2\pm1.9$&$ 0.631\pm0.052$ &\textit{VLA}\\
2009 August 06 &55049.60 & $0.76\pm0.38$&$ 0.191\pm0.046$ &\textit{VLA}\\
&&&&\\
2010 February 13 &55240.01 & $10.6\pm0.9$&$ 0.28\pm0.05$ &\textit{ATCA}\\
\hline  
\hline
\end{tabular}
\small
\\ (1) \textit{VLA} flux density from McClintock et al. (2009); (2) \textit{VLA} flux density from Rupen et al. (2005); (3) \textit{VLA} flux density from Jonker et al. (2010); (4) \textit{Swift} and \textit{Chandra} fluxes
from Jonker et al. (2010) [see section \ref{sec:jonker}]; 
\\ (A) ASM count rate converted into 3-9 keV unabsorbed flux using \textit{WebPimms} [see section \ref{sec:simult}] ;  (B) X-ray flux obtained by interpolation of the PCA lightcurve [see section \ref{sec:simult}]
\normalsize
\end{table*}

The radio/X-ray correlation in BHXBs is usually observed during the canonical hard state where the radio and X-ray emission are assumed to originate from the compact jets and the corona respectively.
To study this connection in detail, we restricted our data-set to observations in the hard state for which the radio spectrum was indicative of optically thick synchrotron emission from compact jets, i.e. a radio spectral index whose lower limit is $\geqslant -0.3$. We also discarded observations that took place following the first radio flare of an outburst, since the compact jets might be disrupted when discrete ejection events take place \citep[see e.g.,][]{corbel04,fender04,fender09a}. 

The selected radio and X-ray fluxes are summarised in Table \ref{table-correl} and the radio/X-ray plot obtained using these filtered data is shown in Fig. \ref{filtre}. Above $\sim 2 \times 10^{-10} \erg$ in the 3--9 keV X-ray band, we note a clear correlation over almost two orders of magnitude in radio flux density. We fit the data with a power-law function of the form $F_{\rm Rad} = k \, F_{\rm X}^b$, where $F_{\rm Rad}$ is the radio flux density at 8.5 GHz (in mJy) and $F_{\rm X}$ is the unabsorbed 3--9 keV X-ray flux (in $\erg$). We obtain a correlation index $b = 1.38 \pm 0.03$ and a normalisation constant $k = 4.43 \times 10^{12}$\,mJy\,erg$^{-1}$s\,cm$^2$. The correlation index $b \sim1.4$ clearly differs from the range of values ($b \sim$ 0.5--0.7) observed for other BHXBs (e.g., GX 339-4, V404 Cyg, XTE J1118+480; see \citealt{corbel03,gallo03,xue07}; \citealt*{corbel08}). Interestingly, it corresponds to the correlation index found for atoll source neutron stars in the island state \citep{migliari06}, a state that shares similar properties with the hard state of BHXBs. We also note that our derived correlation index is not consistent with the one  found by \citet{jonker10} i.e. $b = 0.18 \pm 0.01$. However, this index is obtained by fitting the low luminosity data of the 2008a outburst. Our work suggests that this low value likely reflects a transition phase (see Section \ref{sec:comparison}).

We note however that it is the high flux data from 2003 that mostly constrain the correlation index since it is the brightest outburst observed to date from \h17. Nonetheless, if we exclude the 2003 data from the fitting process, we obtain the following 99.99 per cent confidence interval for the correlation index: $[1.19, 1.65]$. This remain in good agreement with the previous results and seems to favour a constant slope for all outbursts.

\subsection{Comparison with the `standard' radio/X-ray correlation}
\label{sec:comparison}

\begin{figure*}
	 \includegraphics[width=0.8\textwidth]{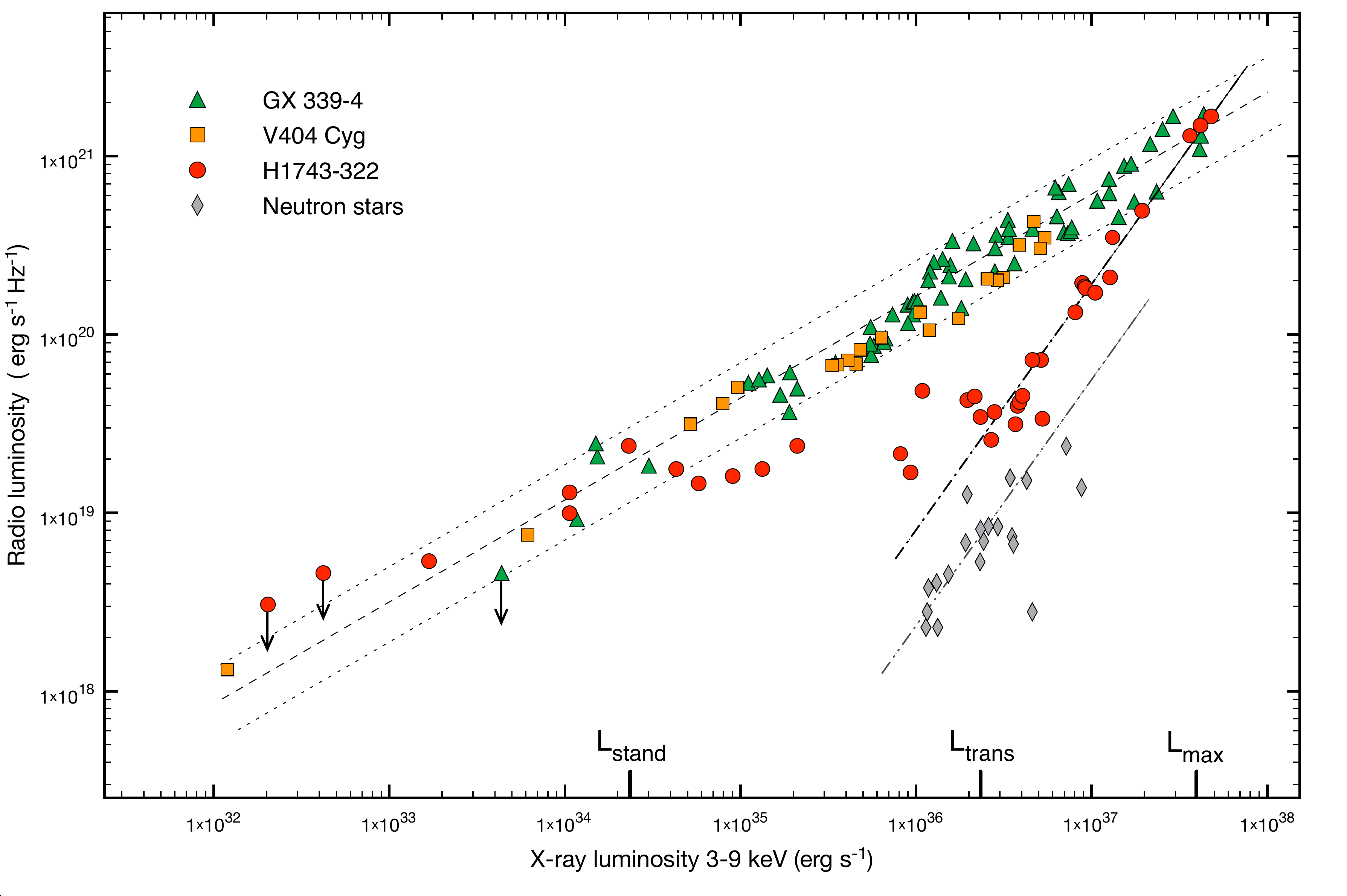} 
	\caption{ 8.5 GHz radio luminosity against 3--9 keV X-ray luminosity for the BHCs \h17, GX 339-4 and V404~Cyg in the hard state. For comparison, we also plot the data from the atoll neutron stars Aql X-1 and 4U 1728-34 in the island state. The dashed line indicates the fit to the GX 339-4 and V404 Cyg data with a relation of the form $L_{\rm Rad} \propto L_{\rm X}^b$ with a correlation index $b \sim 0.6$. The two dotted lines demarcate the dispersion around the correlation. The dashed-dotted and dashed-dotted-dotted lines indicate the fit to the high luminosity data of \h17 and the neutron star data respectively, with a correlation index $b \sim 1.4$. On the x-axis, we also indicate three characteristic X-ray luminosities that will ease the discussion. $L_{\rm stand}$ and $L_{\rm trans}$ are respectively the lower and the upper bound of the transition between the two correlations. $L_{\rm max}$ indicates the point where the steep correlation of \h17 connects to the `standard' correlation at high luminosity. Note that these three luminosities are defined in the 3--100 keV band in the text.}
	\label{RXuniv}
\end{figure*}

To compare the relationship between radio and X-ray fluxes in \h17 with the standard radio/X-ray correlation of black hole and neutron star X-ray binaries, we plot in Fig. \ref{RXuniv} the hard state data from \h17, GX 339-4 (the data cover 7 outbursts over the period 1997--2010 and will be detailed in Corbel et al. [in preparation]), V404 Cyg \citep{gallo03,corbel08} and the atoll neutron stars Aql X-1 \citep{tudose09,miller-jones10} and 4U 1728-34 \citep{migliari06} in the island state. To convert fluxes into luminosity, we used a distance of 8 kpc for GX 339-4 \citep{zdziarski04} and the new distance of 2.39 kpc for V404 Cyg, that was derived using accurate astrometric VLBI observations \citep{miller-jones09a}. For the neutron stars, we used 5.2 kpc for Aql X-1 \citep{jonker04} and 4.6 kpc for 4U 1728-34 \citep{galloway03}.

Fig. \ref{RXuniv} shows that for intermediate luminosities ($\sim 10^{36-37} {\rm \, erg \, s^{-1}}$), \h17 lies significantly below the `standard' correlation for BHXBs but is not compatible either with the neutron star relation. In addition, the figure shows that the deviant points at low luminosity seem to rejoin the standard correlation and then follow it below $2 \times 10^{34}  {\rm \, erg \, s^{-1}}$. The data points between $2 \times 10^{34}  {\rm \, erg \, s^{-1}}$ and $2 \times 10^{36}  {\rm \, erg \, s^{-1}}$ seem thus to reflect a transition between the correlation of slope $b=1.4$ and the standard correlation of slope $b = 0.6$. This supports the idea of a significant change in the coupling between the jets and the corona in \h17 when the source reaches low accretion rates. 

This transition can be crudely fit with a power-law function with index $b = 0.23 \pm 0.07$ between the 3--9 keV luminosities $2 \times 10^{34} \, {\rm erg \; s^{-1}}$ and $2 \times 10^{36} \, {\rm erg \; s^{-1}}$.
The corresponding bolometric (3--100 keV) luminosities in Eddington units are $\, L_{\rm stand} \sim 5 \times 10^{-5} L_{\rm Edd} \, $ and $\, L_{\rm trans} \sim 5 \times 10^{-3} L_{\rm Edd}\, $ for a $10 M_{\sun}$ black hole.

As mentioned in the introduction, \h17 is identified as an outlier of the standard radio/X-ray correlation, as confirmed by our results. Some of the numerous questions associated with this outlier population of sources are whether they follow the same correlation slope as the other BHXBs but with a lower normalisation, and whether they remain below the standard correlation at low and high luminosity.  In the case of \h17, we obtain a correlation index of $b = 1.38 \pm 0.03$. This is the first precise measurement of the radio/X-ray correlation of an outlier. If \h17 is representative of these `radio-quiet' Galactic black holes, our results suggest that their location below the `standard' correlation is a consequence of an intrinsically different coupling between the jets and the corona rather than just a variation of normalisation. In this respect, we point out the recent work by \citet{soleri10} on the outlier Swift J1753.5$-$0127. These authors report a slope of the radio/X-ray correlation of the source lying between 1.0 and 1.4, in good agreement with our results on \h17. Finally, our results suggest that at high ($\sim L_{\rm max}$) and low luminosity ($\leq L_{\rm stand}$), the outliers might not remain below the standard correlation if they follow the same behaviour as \h17 (see Fig. \ref{RXuniv}).

\section{Discussion}

\subsection{Nature of the compact object}
As mentioned above, the mass of the compact object is not yet constrained, so its nature is still uncertain. Based on the radio/X-ray correlation alone and considering the results of \citet{migliari06}, the 1.4 power-law index could suggest that the accreting compact object is a neutron star. However and more tellingly, the overall behaviour of the source during an outburst, and its X-ray spectral and timing features are very similar to other, dynamically confirmed, black hole binaries \citep[e.g., XTE J1550-564;][]{mcclintock09}. The source is thus more likely to be a black hole than a neutron star. In the following discussion, we will therefore consider it as such, but note the caveat that a neutron star primary cannot be entirely ruled out.

\subsection{Radio-quiet or X-ray-loud microquasar?}

As shown in Fig. \ref{RXuniv}, \h17 spans the same range of X-ray and radio luminosity in the hard state as `standard' microquasars. Consequently, should we consider that it displays dimmer radio emission for a given X-ray luminosity or the contrary? In other words, are we facing a radio quiet or an X-ray loud microquasar? In the following, we investigate both hypotheses. First (Section \ref{xloud}), we consider that the outliers have a more radiatively efficient X-ray emitting component than the `standard' microquasars (X-ray loud hypothesis).
 Then (Section \ref{radioquiet}), we consider that the discrepancy between the two populations of sources arise from different jet properties (radiative efficiency, injected power, etc.) leading the outliers to produce fainter radio emission.

\subsection{Radiatively efficient accretion flow in the hard state: the X-ray loud hypothesis}\label{xloud}

We usually define two general classes of accretion flow, depending on whether the gravitational energy of the accreted matter is preferentially released through radiation (radiatively efficient) or carried away with the flow (radiatively inefficient). 

Radiatively efficient flows include for instance, the standard optically thick and geometrically thin accretion disc model \citep{shakura73}, some models of accretion disc coronae (ADC) \citep*[see e.g.,][]{galeev79,haardt91, di-matteo99, merloni02} or the luminous hot accretion flow model \citep[LHAF][]{yuan01}. From simple physical assumptions, the scaling of the X-ray luminosity with accretion rate, in most radiatively efficient flows, is expected to be linear, $L_{X} \propto \dot{M}$. 

Radiatively inefficient flows are expected to produce X-ray emission with $L_{X} \propto \dot{M}^{2-3}$. This is the case of accretion flows dominated by advection in which a significant fraction of the energy is advected instead of radiated away. This advected energy can either cross the event horizon (Advection Dominated Accretion Flow: ADAF; \citealt{ichimaru77,narayan94,abramowicz96}) and/or be expelled in outflows (Advection Dominated Inflow-Outflow Solution: ADIOS; \citealt{blandford99}). In such models, the X-ray emission arises mainly from Compton up-scattering of internal (synchrotron, bremsstrahlung) or external (blackbody emission from outer thin disc) photon fields. A similar relation ($L_{X} \propto \dot{M}^{2-3}$) is also obtained in systems dominated by jet emission, where most of the energy is channeled into the jets \citep{markoff03,markoff05}. The X-rays, in that case, can originate at the base of the jets as optically thin synchrotron emission and/or inverse Compton scattering by the outflowing particles of the jet synchrotron photons (synchrotron self-Compton) and disc photons (external Compton).

We will now show that if we assume the standard emission model of compact jets is valid for the outliers, the steep radio/X-ray correlation we have found implies that the accretion flow is radiatively efficient in the hard state, in contrast to what is usually assumed for BHXBs in this X-ray state.

Let us thus consider the classical assumption stating that the total jet power $Q_{\rm jet}$ is a fraction $f_{j} < 1$ of the (maximal) accretion power $Q_{\rm accr}$:

\begin{equation}
Q_{\rm accr} = \dot{M} c^{2} \qquad \textrm{and} \qquad Q_{\rm jet} = f_{j} \, Q_{\rm accr} \; ,
\label{eq1}
\end{equation}
The fraction $f_{j}$ is usually considered as constant or at least independent of the accretion rate \citep[see e.g.,][]{blandford79, falcke95,heinz03}. $Q_{\rm jet}$ should therefore scale linearly with $\dot{M}$.
Since we restrict ourselves to the standard jet emission model,  we will adopt this assumption. However, there is no strong physical argument justifying that $f_{j}$ is independent of $\dot{M}$, so we shall discuss it in the next section about the radio-quiet hypothesis.

\begin{figure}
	 \includegraphics[width=0.50\textwidth]{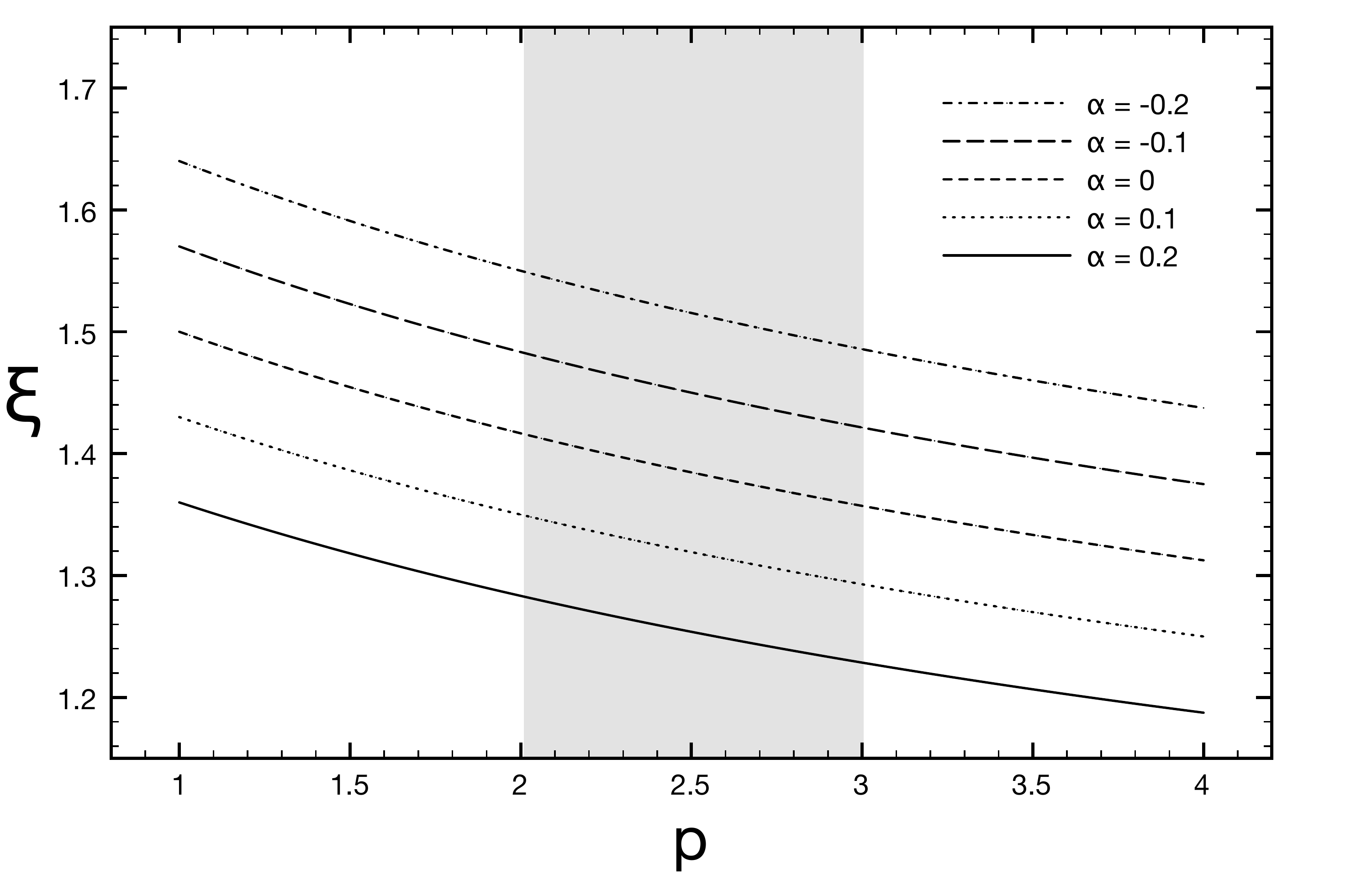} 
	\caption{Variation of the exponent $\xi$ in the relation $L_{\nu} \propto  Q_{jet}^{\xi}$, as a function of the power-law index $p$ of the electron distribution, for several values of the spectral index $\alpha$ of the compact jets. Grey zone delimits the range of values usually assumed for the power-law index $p$.}
	\label{xi}
\end{figure}

From the standard equations for synchrotron emission \citep[e.g.,][]{rybicki79}, one can obtain the following scaling between the jet luminosity $L_{\nu}$ at a given frequency and 
the jet power \citep[see e.g.,][]{heinz03}:

\begin{equation}
L_{\nu} \propto  Q_{\rm jet}^{\xi} \qquad {\rm with} \qquad \xi = \frac{2 p - (p+6)\alpha + 13}{2 (p+4)}
\label{eq2}
\end{equation}
and where $\alpha$ is the spectral index of the jet spectrum (with the convention $L_{\nu} \propto \nu^{\alpha}$). This relation is valid under the assumptions of the standard model, i.e. a conical jet with an initial energy distribution of relativistic electrons in the form of a power-law with index $p$ (for the impact of different assumptions see e.g \citealt{peer09}). 
The classical values assumed for $p$ are in the range $2-3$ \citep[see e.g.,][]{bell78,blandford78,drury82,gallant99,achterberg01}, while the spectral index $\alpha$ of the compact jets in the radio range is usually observed between $-0.2$ and $0.2$. Figure \ref{xi} shows the variation of the exponent $\xi$ as a function of $p$ for different values of $\alpha$. We note that for the fiducial values of $p$ and $\alpha$, $\xi$ varies between 1.22 and 1.58 with an average value of $\sim 1.4$.

If the observing frequency is located in the radio range, Eq. (\ref{eq1}) and Eq. (\ref{eq2}) give $L_{\rm radio} \propto  \dot{M}^{\xi}$.
If we assume that the X-ray luminosity in the hard state can be written in a general way as $L_{X} \propto \dot{M}^{q}$, we expect the 
following relation between radio and X-ray luminosities:

\begin{equation}
L_{\rm radio} \propto  L_{X}^{\xi/q}
\label{eq3}
\end{equation}
Consequently, for $\xi$ between 1.22 and 1.58, the radio/X-ray correlation usually found for microquasars ($L_{\rm radio} \propto  L_{X}^{0.5-0.7}$), requires $q \sim 2-3$ and thus 
a radiatively inefficient accretion flow. On the other hand, the $L_{\rm radio} \propto  L_{X}^{1.4}$ relation of \h17 implies $q \sim 1$, which suggests the X-ray emission in the hard state is produced by a radiatively efficient flow. In this respect, we note that a radiatively efficient flow during the hard state of Cyg X-1 has been found by \citet{malzac09}. Moreover, \citet{rushton10} have shown that GRS 1915+105 could be radiatively efficient during its plateau state. This in turn would be consistent with the fact that it lies on the extension, at high luminosity, of the correlation of \h17 \citep{coriat10}.

As mentioned previously, to obtain the radio/X-ray plot in Fig. \ref{filtre}, we filtered the X-ray data to minimise the thermal contribution of the thin disc. We can thus exclude the possibility that the disc contaminates the 3--9 keV emission enough to produce the observed correlation. 

As far as we know, besides the standard accretion disc solution, the other models of radiatively efficient accretion flows can be divided in two categories according to their geometry:
\begin{enumerate}
\item Hot accretion flow: the standard accretion disc extends to a truncation radius, where it is replaced by a hot and geometrically thick flow in the inner parts.
\item Accretion Disc Corona (ADC): the standard accretion disc extends close to the black hole and is `sandwiched' by a corona of hot plasma.  
\end{enumerate}
In both categories, several models have been developed to explain the properties of BHXBs in the hard state by coupling the accretion flow with steady jets. While the aim of this work is not to review all these models, we briefly examine some examples below.

\subsubsection{Hot accretion flows}

Most hot accretion flow models are found to be radiatively inefficient, at least at low accretion rates (e.g., ADAF, ADIOS, CDAF), and can only exist below a critical accretion rate ($\sim 10^{-1} - 10^{-2} \frac{L_{\rm Edd}}{c^2}$). However, \citet{yuan01} has shown that a hot flow may be maintained above the critical accretion rate when Coulomb coupling between electrons and ions becomes more effective due to increasing density. The obtained accretion flow solution named the Luminous Hot Accretion Flow (LHAF) is described as an extension of the ADAF regime to high accretion rates and is found to be radiatively efficient \citep[see also][]{yuan04,yuan06}. We can thus check if the luminosity range where we observe the steep correlation is compatible with the LHAF regime. We can write the X-ray luminosity of the accretion flow as $L_{X} = \eta \dot{M} c^{2}$ with $\eta$ the efficiency coefficient. $\eta$ should be constant to get $L_X \propto \dot{M}$, which is roughly the case for the LHAF (Yuan private communication). 
Let us thus assume that \h17 is in the LHAF regime in the luminosity range where we observe the correlation of slope 1.4. According to \citet{yuan01}, an ADAF should become a LHAF when $\dot{M}$ exceeds the critical accretion rate $\dot{M}_{\rm crit} \sim 10 \alpha_{v}^2 \dot{M}_{\rm Edd}$ where $\alpha_v$ is the viscosity coefficient and where we have defined the Eddington accretion rate as $\dot{M}_{\rm Edd} = \frac{L_{\rm Edd}}{c^2}$. For the standard value $\alpha_v = 0.1$, we thus have $\dot{M}_{\rm crit} \sim 0.1 \dot{M}_{\rm Edd}$.
Our results show that the steep correlation starts from a 3-100 keV luminosity $L_{\rm trans} \sim 5 \times 10^{-3} L_{\rm Edd}$. Then, for $L_X = L_{\rm trans} = \eta \dot{M}_{\rm trans}c^2$, we have:

\begin{equation}
\eta \dot{M}_{\rm trans} = 5 \times 10^{-3} \mdot_{\rm Edd}
\label{eq4}
\end{equation}

If $\dot{M}_{\rm trans}$ corresponds to the critical accretion rate $\dot{M}_{\rm crit}$ where the LHAF regime starts, then $\eta = 0.05$.
The steep correlation is maintained up to the 3-100 keV luminosity $L_{\rm max} \sim 6 \times 10^{-2} L_{\rm Edd}$ where the intersection with the standard correlation occurs (see Fig. \ref{RXuniv}). If $\eta$ is indeed constant in the LHAF regime, then for $L_X = L_{\rm max}$, the corresponding accretion rate $\dot{M}_{\rm max}$ should be of the order of the Eddington accretion rate. The LHAF hypothesis thus leads to a high but not implausible value of the maximal accretion rate reached in the hard state.

Alternatively, we can estimate $\dot{M}_{\rm trans}$ from the radio luminosity and compare it to the critical accretion rate $\dot{M}_{\rm crit} \sim 0.1 \dot{M}_{\rm Edd}$. Assuming $L_{\rm radio} \propto \mdot^{17/12}$ (equivalent to Eq. (\ref{eq2}) in the case $\alpha = 0$ and $p = 2$), \citet*{kording06b} found an estimate of the normalisation constant between the radio luminosity of the compact jets and the accretion rate:

\begin{equation}
\label{norm-radiojet}
\mdot = \mdot_{0} \left( \frac{L_{\rm radio}}{10^{30}  \rm{\;  erg \; s}^{-1}} \right)^{12/17}  \rm{with \; \;}  \mdot_{0} = 4.0 \times 10^{17} \, \rm{g \; s}^{-1}
\end{equation}

From Fig. \ref{RXuniv}, the radio luminosity corresponding to $L_{\rm trans}$ is $\sim 2.5 \times 10^{19}  \ergs \rm{Hz}^{-1}$. Using Eq. \ref{norm-radiojet} we thus derive $\dot{M}_{\rm trans} \sim 1.5 \times 10^{17} \, \rm{g \; s}^{-1}$. For a $10 \msun$ black hole, this corresponds to $0.1 \, \mdot_{\rm Edd}$, in agreement with the expected accretion rate at the transition to the LHAF regime. The LHAF model is thus a possible explanation for the behaviour of \h17 at high luminosity.

In a similar fashion, hot flow solutions have recently been found (Petrucci private com.) for the Jet Emitting Dic model \citep[JED; see e.g.,][and references therein]{ferreira06,ferreira02,ferreira08,combet08}, in which the flow goes from radiatively inefficient to radiatively efficient as the accretion rate increases. These JED hot solutions have properties very similar to those of one-temperature accretion flow studied by \citet{esin96} in the ADAF regime and revisited for higher accretion rates by \citet{yuan06} in the LHAF regime. The originality of the JED solutions resides in the fact that, by construction, they integrate self-consistently stationary and powerful self collimated jets. 

\subsubsection{Accretion disc corona}

Some ADC models could be also radiatively efficient in the hard state \citep*[see e.g.,][]{galeev79, haardt91, di-matteo99, malzac01, merloni02, merloni03a}. In these models, a fraction $f_{c}$ of the accretion power is dissipated in the corona, likely due to magnetic reconnection, and eventually emerges as X-ray radiation. The X-ray luminosity can be written as $L_{X} \sim f_{c} \dot{M} c^2$. In the case where the coronal plasma is heated by magnetic dissipation, \citet{merloni02} and \citet{merloni03a} have shown that $f_{c}$ is constant when gas pressure dominates in the disc and therefore $L_{X} \propto \dot{M}$ which would explain the steep radio/X-ray correlation. When the radiation pressure dominates in the disc, \citet*{merloni03} have shown that we should also expect a radio/X-ray correlation of the form $L_{\rm radio} \propto L_X^{1.4}$. From the estimates of \citet{merloni03a}, $f_{c}$ should be of the order of 0.02 to 0.07 (for sub-Eddington systems) and would therefore be consistent with the X-ray luminosity range in which we observe the steep correlation. 

From the radio/X-ray correlation point of view, ADC models could be therefore responsible for the X-ray emission in the high luminosity hard state of \h17.  However, the geometry assumed in these models leads naturally to the debate (beyond the scope of this paper) of whether or not the accretion disc extends close to the black hole.

\subsubsection{Efficient to inefficient transition}

When $L_X < L_{\rm trans}$, our results suggest that \h17 undergoes a transition from the steep to the standard correlation. Under the assumptions on jet emission considered in this section, this transition to a relation $L_{\rm radio} \propto L_X^{0.6}$ indicates that the accretion flow becomes radiatively inefficient, with $L_X \propto \mdot^{2-3}$ below $L_{\rm stand}$. If we assume that Eq. (\ref{eq2}) is valid during this transition (i.e. the radio luminosity is a good tracer of the accretion rate), the variation of radio luminosity between $L_{\rm trans}$ and $L_{\rm stand}$ should correspond to a variation of the accretion rate by a factor of $\sim 2$. The corresponding variation of the X-ray luminosity is $L_{\rm trans}/L_{\rm stand} \sim 100$ and this transition occurs on a timescale of 15 days.
If the accretion rates varies little, as suggested by the almost constant level of radio emission during this transition, the radiative efficiency coefficient $\eta$ of the accretion flow should vary significantly (by a factor of $\sim 50$).

A transition from an efficient to an inefficient regime could be interpreted as a transition from an LHAF to an ADAF. In the framework of the ADC models, this could arise from changes in the properties of the corona heating mechanism. These details should be investigated on theoretical grounds.

Rather than being due to changes in the intrinsic properties of the accretion flow (as in the LHAF-ADAF case), the transition could results from the contribution of two emitting components, one inefficient with $L_{X} \propto \mdot^{2-3}$ and the other radiatively efficient with $L_{X} \propto \mdot$. When both components are present, the efficient component dominates the X-ray flux. Below a given \mdot\, it disappears, leaving only the inefficient component. As an illustration, we can point to the work on XTE J1550-564 by \citet{russell10}. The authors demonstrate the possibility that the origin of the X-ray emission evolves throughout the hard state, being alternatively dominated by thermal Comptonization or direct synchrotron emission from the compact jets. We also point out the work by \citet{rodriguez08a,rodriguez08} on GRS 1915+105, where it is reported that two components are present in the hard X-ray spectrum during the plateau state. One of these components appears to be linked to the radio emission while the other is not. This supports the idea that several components coexist and can dominate alternatively the X-ray band during the hard state.  In our case, we could consider that above $L_{\rm trans}$ the X-ray emission is dominated by the contribution of an efficient accretion flow (as , e.g., those mentioned previously). Below $L_{\rm stand}$, the X-rays become dominated, for instance, by the synchrotron or SSC emission from the base of the jets.

\subsubsection{The radio/X-ray diagram of BHXBs under the X-ray loud hypothesis} 

\begin{figure*}
	 \includegraphics[width=0.9\textwidth]{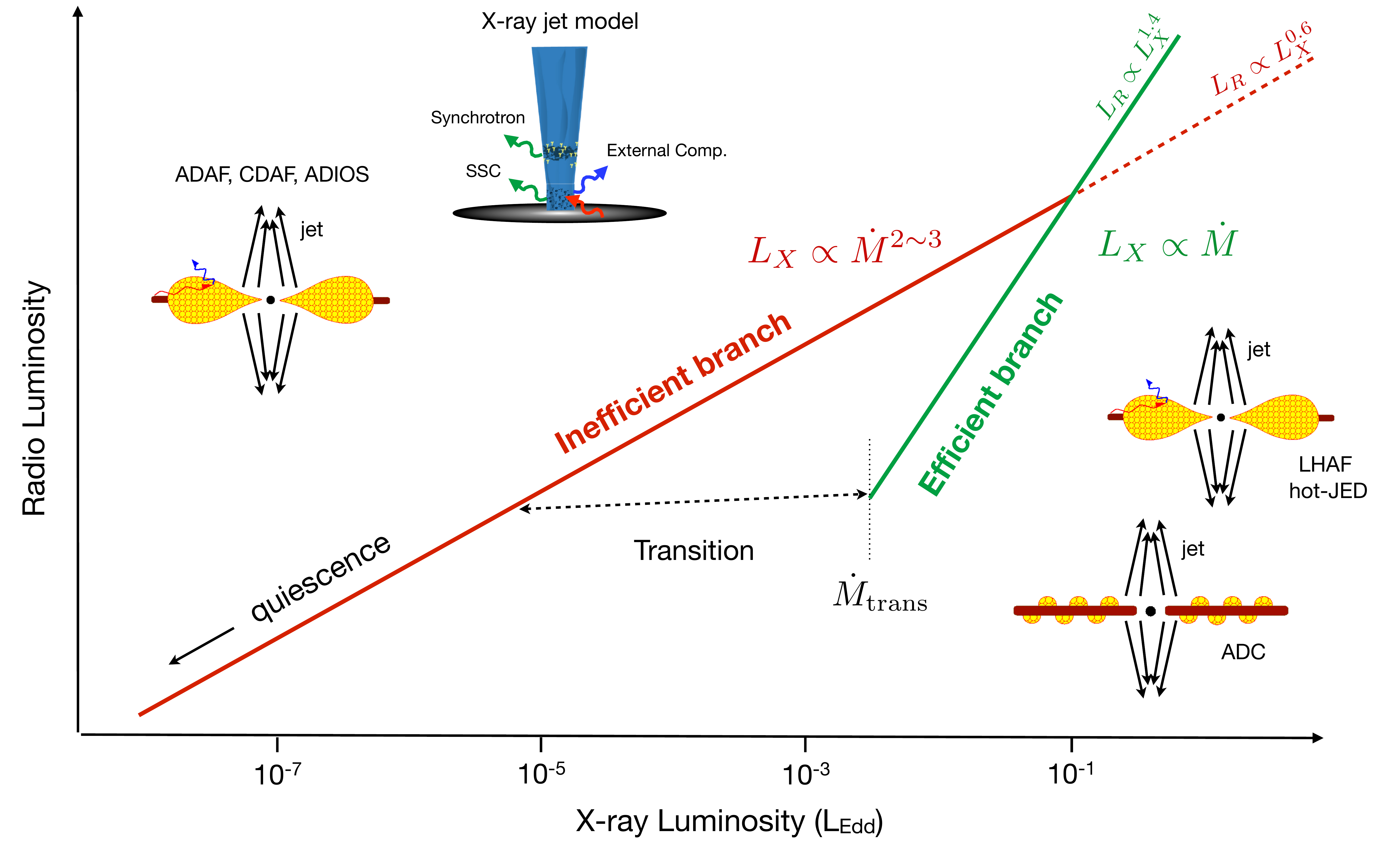} 
	\caption{A schematic drawing of the global radio/X-ray correlation for Galactic black holes. The X-ray luminosity is expressed in terms of the Eddington luminosity for a 10 $M_{\sun}$ black hole. This figure illustrates the case where the correlation $L_{\rm radio} \propto L_X^{1.4}$ of the outliers is a consequence of the coupling between a radiatively efficient accretion flow and a steady compact jet whose radio emission can be described by the relation $L_{\rm radio} \propto \mdot^{1.4}$. We distinguish two branches in this diagram according to the efficiency of the accretion flow and the consequent scaling of the X-ray luminosity with the mass accretion rate. The inefficient branch corresponds to the `standard' radio/X-ray correlation observed for black holes systems like GX 339-4 or V404 Cyg. The inner flow in these systems can be described by, e.g., a hot and inefficient accretion flow such as an ADAF, or by X-ray jet models as proposed by Markoff and collaborators. The efficient branch corresponds the radio/X-ray correlation of the outliers for which the X-ray luminosity in the hard state varies linearly with mass accretion rate. The X-ray emitting component in these systems could be described by , e.g., a hot and efficient accretion flow such as the LHAF or hot-JED solutions  or by some magnetic corona models. We also illustrate on this figure the possibility of a transition between the two branches as suggested by the behaviour of \h17. The pictures representing the different models are adapted from \citet{markoff04} and \citet{nowak04}.}
	\label{sketch}
\end{figure*}

Regardless of the specific models that could explain our results, we can conclude that, as long as the assumptions about the jet physics stated in Eq. (\ref{eq2}) and Eq. (\ref{eq3}) are correct, the accretion flow has to be radiatively efficient (with $L_X \propto \mdot$) above $\sim 5 \times 10^{-3} L_{\rm Edd}$. At lower X-ray luminosities, the radiative efficiency of the accretion flow should decrease significantly to reproduce the transition between the two correlations. Our results suggest then that the X-ray luminosity scales as $\mdot^{2-3}$.
If \h17 is indeed representative of the behaviour of the other outliers, we can thus represent the universal radio/X-ray diagram of BHXBs by the sketch shown in Fig. \ref{sketch}.
This figure summarises the X-ray loud interpretation. We can then distinguish two branches in the radio/X-ray diagram of BHXBs, according to the efficiency of the accretion flow and the consequent scaling of the X-ray luminosity with the mass accretion rate. For a still unknown reason, some BHXBs would remain radiatively inefficient in the hard state up to the transition to the soft state while others (the outliers) would develop a more radiatively efficient accretion flow leading them to follow the efficient branch. We also illustrate the possibility of the transition between branches below the critical accretion rate $\dot{M}_{\rm trans}$.

If this sketch correctly describes the situation, the major issue to address now is to determine which fundamental parameter influences the global evolution of the accretion flow with mass accretion rate and will lead some black hole systems to follow the `efficient' branch and others the `inefficient' branch. In future works we thus need to investigate the influence of parameters such as the orbital period, the environment (e.g., magnetic) of the binary, or perhaps the nature or evolutionary stage of the companion star.
The physical conditions at outer boundary of the accretion disc could also have an influence on the dynamical and radiative structure of the flow \citep{yuan00}.

\subsection{Beyond the standard assumptions on compact jet emission: the radio-quiet hypothesis}\label{radioquiet}

Another way to assess the problem would be to consider that the difference between the standard microquasars and the outliers arises from different jet properties rather than from different accretion flows. 
In this case we can relax the assumptions leading to Eq. (\ref{eq2}) and Eq. (\ref{eq3}). 

We can first consider that the fraction $f_{j}$ (in $Q_{\rm jet} = f_{j} \, Q_{\rm accr}$) of accretion energy injected into the jets is in fact dependent on the accretion rate. For simplicity, we will consider a linear dependence, $f_{j}\propto \dot{M}$. In that case, Eq. (\ref{eq3}) becomes $L_{\rm radio} \propto  L_{X}^{2 \xi/q}$. For a radiatively inefficient accretion flow ($q \sim 2$), this gives the correlation we observe for \h17. 

Whether or not $f_{j}$ is dependent on the accretion rate depends on the details of the jet launching mechanism: e.g., mass loading into the jet, the specific acceleration mechanism or the origin of the magnetic field. A detailed theoretical study is therefore required to address this issue. However, as an example, if we consider the standard theories of magnetically driven jets, we note that the material is accelerated from a given region of the disc. The size of this region is usually considered as constant in the models (and is often the entire disc). But if we assume that, for any reason, this size evolves with accretion rate, it would therefore introduce a dependency of $f_{j}$ on $\dot{M}$. This is however very speculative and we would have to explain why the size of this region changes in some systems and not in others.

Another parameter that could strongly influence the jet emission is of course the strength of the magnetic field embedded in the jet plasma. This will modify the synchrotron emission as a function of the jet power (Eq. \ref{eq2}). \citet{peer09} presented a new model for jet emission in XRBs, in which they showed that the flux at radio wavelengths depends on the value of the magnetic field in a non-trivial way. Above a critical magnetic field strength, the outflowing electrons cool rapidly close to the jet base, leading to a strong suppression of the radio emission. Based on these results, \citet{casella09} proposed that the outliers are sources with magnetic fields above the critical value. With respect to our results, it could also explain the transition phase between the two correlations. If we consider that the magnetic field strength evolves throughout the outburst (e.g., with the accretion rate), the transition could be due to the magnetic field decreasing below the critical value, leading \h17 to the same level of radio emission as GX 339$-$4 and V404 Cyg. However, the model does not explain precisely how the radio luminosity evolves with the injected power and thus with the accretion rate. Therefore, we cannot judge whether or not it is able to reproduce the correlation index of 1.4 we found. We thus encourage further developments of this model.

\section{Conclusions}

In this work we have studied the long term radio/X-ray correlation of the BHC \h17. This source belongs to a group of Galactic black hole X-ray binaries dubbed as outliers of the `universal' radio/X-ray correlation, for being located below the main $L_{\rm Radio} \propto L_{X}^{\sim 0.5-0.7}$ relation. We therefore concentrated our efforts on providing new constraints and improving our understanding of these sources.  Our main conclusions can be summarised as follows:

\begin{enumerate}
\item In the brightest phase of the hard state, we find a tight power-law correlation with a slope $b = 1.38 \pm 0.03$, between the radio flux from the compact jets and the X-ray emission from the inner flow. This correlation is much steeper than usually found for black hole X-ray binaries and is the first precise measurement for an outlier. 
\item When the source reaches a luminosity below $\sim 5 \times 10^{-3} L_{\rm Edd} \, (M/10 M_{\sun})^{-1}$, we found evidence of a transition from the steep $b \sim 1.4$ relation to the standard $b \sim 0.6$ correlation seen in, e.g., V404 Cyg and GX 339$-$4.
\item Additionally, we find that \h17 provides the best constraint to date (with a jet quenching factor of $\sim 700$) supporting the idea of jet suppression during the soft state.
\end{enumerate}
From these results, we discuss several hypotheses that could explain the correlation index along with the transition toward the standard correlation:
\begin{enumerate}
\item We first show that if the standard scaling, $L_{\rm radio} \propto \dot{M}^{1.4}$, between the jet radio emission and the accretion rate is valid, then our results require a radiatively efficient accretion flow that produces the X-ray emission in the hard state at high accretion rate. Ultimately, the flow has to become radiatively inefficient below a critical accretion rate, to account for the transition.
\item We also investigate the possibility that our results arise from the outflow properties of the source rather than from the accretion flow. We show in particular that if we relax the assumption that the jet power is a fixed fraction of the accretion power and we consider this fraction linearly dependent on the accretion rate, we can obtain the required correlation with an inefficient accretion flow.
\end{enumerate}
Further investigations are now needed, to determine which fundamental parameter of the binary systems or their environments, can lead BHXBs of similar appearance to develop different accretion or ejection flows. 

\section*{Acknowledgments}

MC and SC would like to thank Julien Malzac, Pierre-Olivier Petrucci, Elmar K\"ording, Sera Markoff, Feng Yuan and Michiel van der Klis for useful comments and discussions and Philip Edwards for prompt scheduling of the ATCA observations. The data on the 2009 outburst were collected by the JACPOT XRB collaboration \citep{miller-jones10a}, as part of an ongoing VLBA large project.  In particular, the authors would like to acknowledge the roles played by Ron Remillard, Michael Rupen and Vivek Dhawan in this effort.
The authors would also like to thank the anonymous referee for his critical reading that helped to improve the style and content of this paper. 

The research leading to these results has received partial funding from the European CommunityÕs Seventh Framework Programme (FP7/2007-2013) under grant agreement number ITN 215212 ÓBlack Hole UniverseÓ.
This research has made use of data obtained from the High Energy Astrophysics Science Archive Research Center (HEASARC), provided by NASA's Goddard Space Flight Center. The ATCA is part of the Australia Telescope funded by the Commonwealth of Australia for operation as a National Facility managed by CSIRO. The National Radio Astronomy Observatory is a facility of the National Science Foundation operated under cooperative agreement by Associated Universities, Inc.

\bibliographystyle{mn2e}
\bibliography{biblio}

\bsp

\label{lastpage}

\end{document}